\documentclass[a4paper,12pt]{iopart}
\usepackage{iopams}
\usepackage{endnotes}

\usepackage{graphics}
\usepackage{latexsym}
\usepackage{epsfig}
\usepackage{amsbsy}

\begin{document}

\title{Perturbative corrections	to photon coincidence spectroscopy}

\author{L.\ Horvath
	and B.\ C.\ Sanders}

\address{Department of Physics, Macquarie University,
Sydney, New South Wales 2109, Australia}

\date{Received: 2000 / Revised version: } 

\begin{abstract} 
Photon coincidence spectroscopy is a promising technique 
for probing the nonlinear regime of cavity quantum electrodynamics
in the optical domain, however its
accuracy is mitigated by two factors:
higher--order photon correlations, which contribute to an enhanced
pair count rate, and non--simultaneity of emitted photon pairs from   
the optical cavity.
We show that the technique of
photon coincidence spectroscopy is effective in the presence of
these effects if the quantitative predictions  
are adjusted to include non--simultaneity 
and higher--order correlations. 
\end{abstract}


\maketitle 

\section{Introduction}
\label{intro}

Photon coincidence spectroscopy (PCS) provides an accurate means for
detecting quantum field effects in cavity quantum electrodynamics (CQED) 
(Carmichael {\em et~al.}~1996, Sanders {\em et~al.}~1997,
Horvath {\em et al.}~1999, Horvath and Sanders~2001).
In applications of CQED to quantum 
information, where entanglement between the cavity field and the
internal degrees of freedom of the atom passing through the cavity is
central to the scheme 
(Turchette {\em et~al.}~1995), PCS offers an accurate method for 
probing the quantum field in the cavity.  In two--photon coincidence 
spectroscopy (2PCS), the atom--cavity system is driven by a bichromatic 
field (as for nonlinear spectroscopy; see Thompson {\em et~al.} 1998), 
and a detector measures the two--photon coincidence rate (2PCR)
in the cavity output field.  Two--photon spectral peaks are expected
for particular sum frequencies of the bichromatic driving field (one
frequency is fixed and the other scanned across some range of frequencies).
Certain two--photon spectral peaks would occur only if the quantum 
description of the intracavity field is valid and not occur if the
intracavity field can be described by a semiclassical field theory.

In order to test the quantum description of the intracavity field,
accuracy in predicting the locations of the two--photon spectral peaks
is very important. The height or the width of the two--photon spectral
peaks does not need to be determined by theory accurately, but the
centre of the peak, as a function of scanning field frequency, is
important.  We consider here two effects on the shifts of these peaks:
higher--order photon correlations that contribute to the 2PCR and
the non--simultaneity of the photon pairs as they exit the cavity.
These two cases contribute to small shifts of the 2PCR peaks but must
be accounted for in establishing accurate predictions by 2PCS.

\section{Master equation and two--photon count rates}
\label{sec:Master:Twophoton}

In the electric--dipole and
rotating--wave approximations, the Jaynes--Cummings (JC) Hamiltonian 
for the two--level atom 
(2LA) coupled to a single mode is (Jaynes and Cummings~1963)
\begin{equation} 
\label{JCH} 
H(g) = \omega (\sigma_z + a^{\dagger} a)
	+ i g (a^{\dagger} \sigma_- - a \sigma_+ ), 
\end{equation}
with~$g$ a position--dependent dipole coupling strength, 
$a$ and $a^{\dagger}$ the annihilation and creation operators 
for photons in the cavity field, 
$\sigma_+$, $\sigma_-$, and $\sigma_z$ the 2LA  
raising, lowering and inversion operators, respectively, 
and $\hbar = 1$. Single--atom cavity QED in the optical domain
(Hood {\em et~al.}~2000,  Pinkse {\em et~al.}~2000) 
validates the single--atom approximation of Eq.~(\ref{JCH}).
Provided that the atoms move sufficiently slowly through the cavity 
(Carmichael {\em et~al.}~1996, Sanders {\em et~al.}~1997),
and the position is randomly varying,   
the coupling strength~$g$ is also random:
hence, a coupling strength distribution~$P(g)$ can be  
constructed
(Sanders {\em et~al.}~1997), and we assume the~$P(g)$ depicted in 
Fig.~5 of Sanders {\em et~al.}~1997,
which applies for a single TEM$_{00}$
cavity mode and atoms travelling transverse to the cavity axis.
The atoms first pass through a rectangular mask, centred at an
antinode with size~$w_0\times\lambda/10$ (with~$w_0$ the cavity mode
waist and~$\lambda$ the optical wavelength). The atoms then
traverse the cavity after passing through the mask.
We restrict~$Fg_{\rm max} < g < g_{\rm max}$ for~$g_{\rm max}$
the coupling strength at an antinode along the cavity longitudinal axis 
and~$F$ an effective cut--off term. 

\begin{figure}[1]
\begin{picture}(200,400)(0,0)
\font\gnuplot=cmr10 at 10pt
\gnuplot
{\resizebox{320pt}{370pt}{\includegraphics{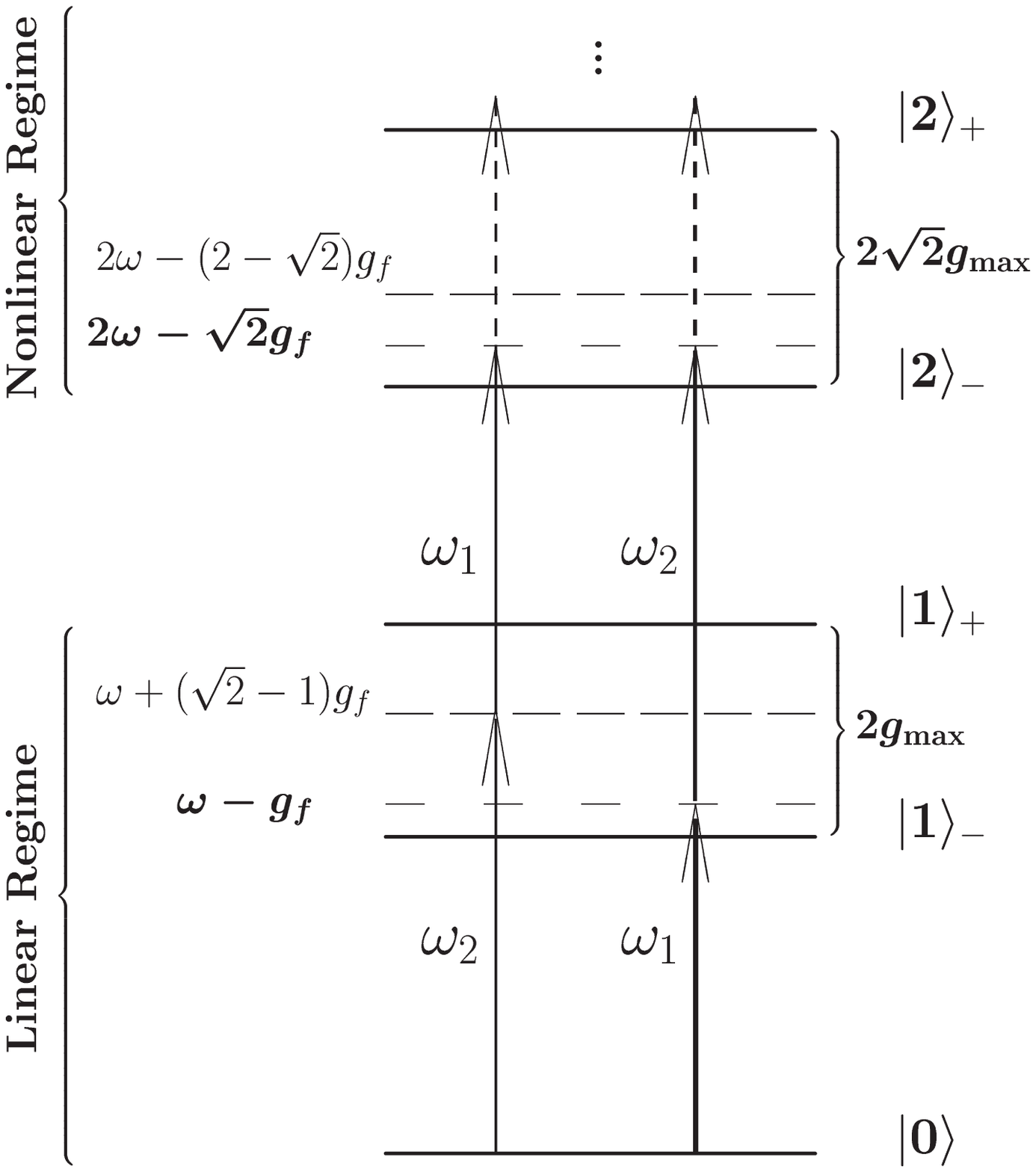}}}
\end{picture}
\caption{Two--photon excitation scheme from the ground state 
	$\left\vert 0 \right\rangle$ to the first two excited couplets
	$\left\vert n \right\rangle_{\varepsilon}$ 
	($n\in\{1,2\}$, $\varepsilon \in \{-,+\}$) of the dressed states.
	The inhomogeneous broadening of the 
	couplets~$\left\vert 1 \right\rangle_\varepsilon$ 
	and~$\left\vert 2 \right\rangle_\varepsilon$ 
	is~$2\hbar g_{\rm max}$ and~$2\sqrt2 \hbar g_{\rm max}$, respectively. 
	Two two--photon excitations to the second couplet are
	depicted for a bichromatic driving field with one component
	of amplitude~${\cal E}_1$ and the other with amplitude~${\cal E}_2$.
	The excitation pathway on the right ($\omega_1$ then~$\omega_2$)
	excites resonantly from~$\vert 0\rangle$ to~$\vert 1\rangle_-$
	and then may excite resonantly to either~$\vert 2\rangle_{\pm}$.
	The excitation pathway on the left ($\omega_2$ then~$\omega_1$)
	excites resonantly from~$\vert 0\rangle$ to~$\vert 1\rangle_+$
	to~$\vert 2\rangle_-$ for~$g=(\sqrt2-1)g_f$.}
\label{fig:ladder}
\end{figure}

The JC spectrum for the Hamiltonian (\ref{JCH}) 
is depicted in Fig.~\ref{fig:ladder}
and the
 `dressed states' of the atom--cavity system
are designated as the lowest--energy 
state~$|0\rangle \equiv \vert 0 \rangle_{\rm cav}
	\otimes \vert {\tt g} \rangle_{\rm atom} 
	\equiv \vert 0, {\tt g} \rangle$,
and, for $n$ a positive integer, the `excited' 
state couplets~$|n \rangle_{\pm} \equiv i/\sqrt{2}
	 \left( \left\vert n-1, {\tt e} \right\rangle
	 \pm i \left\vert n, {\tt g} \right\rangle \right)$,
with~$ |n\rangle $ the Fock state of the cavity mode and
$ |{\tt g}\rangle \, ( |{\tt e}\rangle )$
the ground (excited) state of the 2LA. 
Averaging over~$P(g)$ yields inhomogeneous spectral 
broadening (due to atomic position variability).

Two--photon excitation is provided by driving the atom directly, as it
traverses the cavity, 
with a bichromatic field~${\cal E}(t) = {\cal E}_1 e^{-i\omega_1 t} 
	+ {\cal E}_2 e^{-i\omega_2 t}$.
The driving--field frequency~$\omega_1$ is fixed and resonantly 
excites~$|0\rangle\longleftrightarrow |1\rangle_-$  
for the subensemble~$g=g_f=\omega-\omega_1$ 
corresponding to~$P(g)=\delta(g-g_f)$.
The scanning--field frequency $\omega_2$
excites~$\vert 1\rangle_-\longleftrightarrow\vert 2\rangle_{\pm}$
for~$P(g)=\delta(g-g_f)$.
Enhanced rates of photon pair detection are then sought as the
scanning frequency~$\omega_2$ is varied 
such that~$\omega_1+\omega_2$ is resonant
with some transition~$\vert 0\rangle\longleftrightarrow \vert 2\rangle_{\pm}$
as depicted in Fig.~\ref{fig:ladder}.

The master equation for this system
(Sanders {\em et~al.}~1997)
can be expressed as
$\dot{\rho} = {\cal L}\rho$ for 
${\cal L}={\cal L}_{\rm eff}+{\cal D}+{\cal J}$, i.e.~a 
sum of a Liouville operator~${\cal L}_{\rm eff}$,
an explicit time--dependent Liouville operator ${\cal D}$ and a `jump' 
term~${\cal J}$.
We introduce $\delta \equiv \omega_2-\omega_1$
and work in the rotating picture with respect to the driving--field
component~$\omega_1$.

For $\Xi(g)=ig(a^{\dagger}\sigma_- -a\sigma_+)$, the
effective Hamiltonian is 
\begin{eqnarray}
\label{eq:Heff}
H_{\rm eff}(g,{\cal E}_1)
	&=& \left( \omega - \omega_1 \right) (\sigma_z + a^{\dagger} a) 
	+ \Xi(g)+\Upsilon({\cal E}_1)	\nonumber	\\
	&&- i\kappa a^{\dagger} a - i(\gamma/2) \sigma_+ \sigma_-,
\end{eqnarray}
with~$2\kappa$ the decay rate for the cavity,~$\gamma$ the
inhibited spontaneous emission rate for the partially confined atom
passing through the cavity,
$\Upsilon({\cal E}_1)=i{\cal E}_1 (\sigma_+-\sigma_-)$ a monochromatic 2LA
driving
term, and
\begin{equation}
\label{eq:Leff}
{\cal L}_{\rm eff} (g, {\cal E}_1) \rho
	= -i \left[ H_{\rm eff}(g, {\cal E}_1) \rho 
		- \rho H_{\rm eff}^{\dagger} (g, {\cal E}_1) \right] .
\end{equation}
The jump term superoperator $\cal J$ is given by
${\cal J} \rho = 2 \kappa a \rho a^{\dagger}+\gamma \sigma_- \rho \sigma_+$,
and 
${\cal D} \rho
	= -i \left[ \Upsilon({\cal E}_2 e^{-i\delta t}),\rho \right]$.

Solving the master equation for~$\dot{\rho}$, and accounting
for atomic position variability, yields a solution
\begin{equation}
\label{density:matrix}
\bar{\rho} \equiv \int_{Fg_{\rm max}}^{g_{\rm max}} P(g) \rho(g) dg. 
\end{equation}
For a bichromatic driving field, the density 
matrix does not settle to a steady state value, but, 
in the long--time limit~$t \longrightarrow \infty$, the
Bloch function expansion is
\begin{equation}
\label{Bloch}
\lim_{t\rightarrow \infty} 
\bar{\rho}(t) = \sum_{m=0}^{\infty} \bar{\rho}_m e^{i m \delta t},
\end{equation}
with
$\overline{\rho}_m$
time--independent matrices.
As the photocount integration time is expected to be long compared to the 
detuning~$\delta$, it is reasonable to assume that rapidly oscillating
terms average out and therefore
approximate $\rho(g)$
by truncating the expansion~(\ref{Bloch}). 

An experimental signature for entanglement is the 2PCR,
which is given by the expectation value 
\begin{equation}
\label{eq:w2n}
w^{(2)}(t,\delta, {\cal E}_1)
	= (2\kappa)^2\left\langle : n(t_0) n(t_0+t) : \right\rangle,
\end{equation} 
for 
$n(t) \equiv a^{\dagger}(t) a(t)$
and `$: \, :$' referring to normal ordering. The coefficient~$(2\kappa)^2$
accounts for the photon pair rate of emission from the cavity.
The expectation value corresponds to a trace over~$\overline{\rho}$ in
Eq.~(\ref{density:matrix}) for some choice of~$P(g)$.
The 2PCR is of interest in the asymptotically large time~$t_0$ limit,
where transients disappear and $t_0$ is unimportant: hence the~$t_0$ 
dependence on the left--hand side of Eq.~(\ref{eq:w2n}) is ignored.
The 2PCR is obtained by collecting photons over some duration,
which we call the window time~$\tau_{\rm w}$.
Thus, the actual 2PCR is an integral, or a double integral,
over~$t$ of $w^{(2)}(t,\delta,g)$, yielding an effective
two--photon coincidence rate of $w^{(2)}(\delta,g)$
for a window time of $\tau_{\rm w}$.
The details of the window time, and the choice of performing 
a single or a double integral, are fully discussed in 
Sec.\ {\ref{sec:nonsimultaneity}}.
For this Section, we assume that the window time is sufficiently
small that
(Sanders {\em et~al.}~1997)
\begin{equation}
\label{eq:w2}
w^{(2)}(\delta,{\cal E}_1) = (2\kappa)^2\langle : n^2 : \rangle.
\end{equation}

For 2PCS, the major obstacle to detect entanglement 
is the background 2PCR due to
two photons of angular frequency~$\omega_2$ contributing
to excitation to the second couplet, generally by off--resonant,
but nonnegligible, transitions.
The method for overcoming this problem is called `background subtraction'.
In this technique, the experiment is performed twice,
once with a bichromatic driving field to 
obtain~$w^{(2)}(\delta,{\cal E}_1)$
and again but with a monochromatic driving field at frequency~$\omega_2$
(i.e.\ ${\cal E}_1=0$).
The difference between the two 2PCRs, for the cases that
${\cal E}_1 \neq 0$ and ${\cal E}_1 = 0$, is given by
\begin{equation}
\Delta^{(2)}(\delta,{\cal E}_1) = w^{(2)}(\delta,{\cal E}_1)
	- w^{(2)}(\delta,{\cal E}_1=0) ,
\end{equation}
which we refer to as the `difference--2PCR'. The 2PCR with and
without background subtraction is depicted in Fig.~\ref{fig:Fig1},
and the importance of background subtraction is apparent for the
two--photon spectral peaks in the
domain~$\vert \tilde\delta\vert\leq1$ 
for $\tilde\delta\equiv (\omega_2-\omega)/(\omega-\omega_1)$
and~$P(g)=\delta(g-g_f)$.

\section{Multi--photon contributions}
\label{sec:multiphoton}

The emphasis thus far has been on 2PCS, with its two--photon
decay process leading to an enhanced 2PCR.
Although two--photon events dominate the dynamics,
multi--photon effects cannot be ignored. In fact, multiphoton contributions
can be used to extract signatures of transitions from higher--order couplets
in the JC ladder. This method is known as 
multi--photon coincidence spectroscopy
(Horvath {\em et~al.}~1999).

In principle multi--photon emissions from the cavity can 
be distinguished from two--photon emissions, but, in practice, a
photodetection system located outside the cavity, which is designed
to detect two coincident photons passing through the cavity
mirror, cannot distinguish between
two coincident photons and the rarer case of $N$
coincident photons with $N>2$.
In addition to the case of multi--photon emissions 
in the cavity output field, there are also cases where
multiple photons simultaneously exit the cavity mirror
{\em and} exit the side of the cavity.
Hence, the two--photon coincidence rate includes contributions from 
multiple photons emitted from the cavity as well as two
photons emitted by the 
cavity and at least one photon leaving the side of the cavity.

\subsection{Correlation Functions}
\label{subsec:correlationfunctions}

In order to treat multiphoton emissions, the correlation 
function~$w^{(2)}(t,\delta,{\cal E}_1)$ in Eq.~(\ref{eq:w2}),
must be generalized to the multiphoton 
case~$w^{(\ell)}(\mbox{\boldmath{$t$}},\delta,{\cal E}_1) $,
for $\ell \ge 2$ the number of photons emitted.
These photoemissions take place at random times
\begin{equation}
\label{eq:photemtimes}
\mbox{\boldmath{$t$}}=(t_1, t_2, \ldots, t_{\ell-1}),
\end{equation}
with~$\mbox{\boldmath{$t$}}$ a vector consisting of the photoemission times.
Only $\ell-1$ photoemission times are required for~$\ell$ 
photoemissions,
as each photoemission time is determined with 
respect to the time
of the first photoemission which can be fixed at $t_0 = 0$.
Thus, the  higher contributions to the 2PCR are given by  
$\sum_{\ell = 2}^{\infty} w^{(\ell)}(\mbox{\boldmath{$t$}},
\delta, {\cal E}_1)$.

A compact notation for $w^{(\ell)}$
is possible if we introduce a single notation for
the number operator~$n$ corresponding to the cavity field
and for the atomic inversion operator~$\sigma_+ \sigma_-$.
We introduce the operator $\Theta_i$, with $i \in \{ 0,1 \}$,
and $\Theta_0 \equiv n$, $\Theta_1 \equiv \sigma_+ \sigma_-$.
By introducing the length $\ell-2$ vector
$\mbox{\boldmath{$k$}}$, such that each $k_i \in \{ 0,1 \}$ $\forall i$,
we obtain the compact expression
\begin{eqnarray}
\label{eq:multiphoton}
w^{(\ell)}(\mbox{\boldmath{$t$}},\delta, {\cal E}_1)
	\equiv \sum_{\mbox{\boldmath{$k$}}}
	\alpha_{\mbox{\boldmath{$k$}}} \left\langle : n \cdot n(t_1)
		\cdot \prod_{s=2}^{\ell-1}
		\Theta_{k_{s-1}}(t_s) : \right\rangle,
\end{eqnarray}
with~$\{\alpha_{\mbox{\boldmath{$k$}}}\}$ a set of real constant coefficients. 
The sum is over all~$2^{\ell-2}$
distinct vectors
\begin{eqnarray}
\label{eq:kvec}
\mbox{\boldmath{$k$}}\in && \left\{(0,0,\ldots,0,0,0) \right.,
(0,0,\ldots,0,0,1), \nonumber \\
&& (0,0,\ldots,0,1,0),
\left. \ldots,(1,1,\ldots,1,1,1)\right\}.
\end{eqnarray}
The
operators~$n$ and~$\sigma_+\sigma_-$ correspond to quantum numbers 
{\em inside}
the cavity: the corresponding extra--cavity terms are~$2\kappa n$ 
and~$\gamma\sigma_+\sigma_-$. These coefficients correspond to the
rates for the quantum to leave through the cavity mirror
and through the side of the cavity, respectively. 
For each~$n$,
we introduce the multiplicative factor of~$2\kappa$, and the 
coefficient~$\gamma$ applies for each~$\sigma_+\sigma_-$ in the 
product. Thus,
\begin{equation}
\label{eq:alpha}
\alpha_{\mbox{\boldmath{$k$}}}=(2\kappa)^{\ell-K}\,\gamma^K
\; \mbox{for} \;
	K\equiv\sum_{i=1}^{\ell-2} k_i.
\end{equation}
For example, let us consider~$\ell=4$. In this case
\begin{eqnarray}
\label{eq:w5}
w^{(4)}((t_1,t_2,t_3),\delta,{\cal E}_1)&=&
(2\kappa)^2\left\langle : n\cdot n(t_1)\cdot [ (2\kappa)^2
\Theta_0(t_2) \Theta_0(t_3) \right.
\nonumber \\
&&+ 2\kappa\gamma\Theta_0(t_2)\Theta_1(t_3)
+ 2\kappa\gamma\Theta_1(t_2)\Theta_0(t_3) \nonumber \\
&&+ \gamma^2\left. \Theta_1(t_2)\Theta_1(t_3) ]:\right\rangle.
\end{eqnarray}

We observe that $w^{(\ell)}$ is the correlation function for two photons
at the initial time $t_0$ and a later time
$t_1$, plus the set of all photons emitted from the cavity
or out the side at the other times $\{ t_2,t_3,\ldots,t_{\ell-1}\}$.

In Sec.\ {\ref{sec:nonsimultaneity}} we consider the window time
$\tau_{\rm w}$ in detail, but here we make the small window time
assumption~$\kappa\tau_{\rm w}\ll 1$.
Hence the $\mbox{\boldmath{$t$}}$ dependence is negligible, and
\begin{eqnarray}
\label{eq:pertn}
w^{(n \ge 2)}({\cal E}_1,\delta) = \sum_{m=0}^{n-2} (2\kappa)^{n-m}\,\gamma^m
\langle\sigma_+^m \hat{a}^{\dag\,n-m} \hat{a}^{n-m}
	\sigma_-^m\rangle.
\end{eqnarray}
Expression~(\ref{eq:pertn}) is valid provided that the
window time is small, as we assume in this Section.

\subsection{Three--photon coincidence rate}
\label{subsec:threephotonrate}


In Fig.~\ref{fig:three} the three--photon coincidence rate (3PCR) is shown
for the cases~$\left\langle a^{\dag\,3}a^3\right\rangle$ 
and~$\left\langle \sigma_- a^{\dag\,2}a^2\sigma_+\right\rangle$ with 
fixed coupling strength~$g=g_f=9\kappa$. Vertical lines depict the seven 
values of~$\tilde\delta$ which would lead to on-- or off--resonant
excitation pathways which, after the third photon absorption, are 
on--resonance with either~$\vert3\rangle_-$ or~$\vert 3\rangle_+$.
These vertical lines serve as a guide to where peaks in the 3PCR
may be expected. The values of~$\tilde\delta$ for which these
vertical lines occur are given by the resonance 
conditions
\begin{eqnarray}
\label{mult:res1}
	3\omega -\sqrt3 g &=& 3\omega_1, 	\\
\label{mult:res2}
	3\omega\pm\sqrt3 g &=& 2\omega_1+\omega_2, \,
	\tilde\delta=2\pm\sqrt{3} g/g_f ,	\\
\label{mult:res3}
        3\omega\pm\sqrt3 g &=& \omega_1+2\omega_2, \,
	\tilde\delta=\left(g_f\pm\sqrt{3}g \right)/2g_f ,  \\
\label{mult:res4}
        3 \omega\pm\sqrt3 g &=& 3\omega_2,\,
	\tilde\delta = \pm g/\sqrt{3}g_f.
\end{eqnarray}
The 3PCR peaks are 
observed near these
vertical lines, although small shifts of the peaks are evident. 
The reason for these peak shifts is the influence of 
competition between excitation pathway,
as discussed in 
Horvath and Sanders 2001
for 2PCS. These shifts in peak
positions are small and do not affect the analysis here.
The total 
3PCR is the weighted sum of the full and dotted curves of 
Fig.~\ref{fig:three}:
$(2\kappa)^2(2\kappa\left\langle a^{\dag\,3}a^3
\right\rangle + \gamma \left\langle \sigma_+ a^{\dag\,2} a^2 \sigma_-
\right\rangle)$.

\subsection{Analysis of the three--photon count rate}
\label{subsec:3PCR}

In order to understand the detailed structure of Fig.~\ref{fig:three},
we modify the Liouvillean superoperator~${\cal L}$
by artificially eliminating particular driving terms 
responsible for certain peaks. 

Although a thorough examination of the peak structure in Fig.~\ref{fig:three}
requires an analysis of both rates~$\left\langle a^{\dag\,3}a^3 \right\rangle$ 
and~~$\left\langle \sigma_+a^{\dag\,2}a^2 \sigma_-\right\rangle$,
an analysis of one of the two graphs is sufficient.
The rationale for the peak structure is readily extended to explain the 
similar peak structure observed 
for the other rate. Without loss of generality, we 
choose, in this Section, to study in detail the peak structure for the
3PCR~$\left\langle a^{\dag\,3}a^3 \right\rangle$. 
This analysis of the 3PCR structure is important in 
Subsection~\ref{subsec:difftwoph}, as the 3PCR is responsible
for perturbative corrections to the desired 2PCR, which provides
the experimental signature of quantum field effects. In addition,
the 3PCR structure could be used to identify further signatures
of quantum field effects compared to the 2PCR signature alone. 

The isolation of specific transitions is obtained by the following 
procedure. For example, let us consider the influence on the 3PCR of 
the~$\left\vert 0 \right\rangle \longleftrightarrow \left\vert 1 \right\rangle_-$
transition. We can write the effective Hamiltonian~(\ref{eq:Heff}) as 
a matrix in the dressed--state basis. To isolate this influence, we can
set the matrix elements~$\left\langle 0 \right \vert \Upsilon({\cal E}_1)
\left\vert 1 \right\rangle_-$ and~$\left.\right._-\!\left\langle 1 
\right \vert \Upsilon({\cal E}_1)\left\vert 0 \right\rangle$ to be zero 
where~$\Upsilon({\cal E}_1)$ is the driving term of Eq.~(\ref{eq:Heff}).
In addition
the matrix elements of the driving term~$\left\langle 0
\right\vert \Upsilon({\cal E}_2\exp(-i\delta t)) 
\left\vert 1 \right\rangle_-$ and its complex conjugate can
both be set to zero. The jump term~${\cal J}$ is not modified because
only the driving terms and their effects are of concern in this analysis. 


The result of setting the matrix element~$\left\langle 0
\right \vert \Upsilon({\cal E}_1)
\left \vert 1 \right\rangle_-$ and its
complex conjugate to zero for calculating the 
rate~$\left\langle a^{\dag\,3}a^3 \right\rangle$ 
is depicted in Fig.~\ref{app:3PCR:1}(a). 
Similarly Fig.\ 
\ref{app:3PCR:1}(b)
shows the graph for~$\left\langle 0
\right \vert \Upsilon({\cal E}_2\exp(-i\delta t))
\left \vert 1 \right\rangle_-$ and its conjugate set to zero.
We observe that eliminating these two matrix 
elements causes a dramatic reduction of neighbouring peaks at 
both~$\tilde\delta=-\left(\sqrt2-1\right)$, which is depicted as peak~$ii$,
and~$\tilde\delta=-\left(\sqrt3-1\right)/2$, which is depicted as peak~$iii$
in Fig.~\ref{fig:three}.
The reason for this reduction is that~$\omega_1$ corresponds to 
the~$\left\vert 0 \right\rangle\longleftrightarrow 
\left\vert 1 \right\rangle_-$
transition.  
To understand the next step we refer to Fig.~\ref{app:3PCR:2}(b)
which depicts~$\left\langle a^{\dag\,3}a^3 \right\rangle$ 
with~${\cal E}_2=0$ for 
the~$\left\vert 1\right\rangle_- \longleftrightarrow\left\vert 2 \right\rangle_-$
transition. In this figure it is clear that the photon of frequency~$\omega_2$
dominates the~$\left\vert 1\right\rangle_- \longleftrightarrow\left\vert 2 
\right\rangle_-$ transition. For the peak~$\tilde\delta=-(\sqrt2-1)$, this 
transition is on
resonance, but, for the peak~$\tilde\delta=-(\sqrt3-1)/2$, the 
transition is slightly off--resonant.
 

The final contribution to the rate appears in Fig.~\ref{app:3PCR:3}(b)
where the contribution of an~$\omega_2$ photon to the peaks 
is apparent. 
However, the transition~$\vert 2\rangle_-\longleftrightarrow\vert 3 \rangle_-$
for the peak~$\tilde\delta=-(\sqrt2-1)$ is not resonant, but the 
transition is resonant for the peak at~$\tilde\delta=-(\sqrt3-1)/2$. 
From the three figures~\ref{app:3PCR:1}(a),
\ref{app:3PCR:2}(b) and \ref{app:3PCR:3}(b), we observe that the peak rate
at~$\tilde\delta=-(\sqrt2-1)$ and~$\tilde\delta=-(\sqrt3-1)/2$
is overwhelmingly due to a sequence of 
an~$\omega_1$ photon followed by two~$\omega_2$ photon absorptions.
By removing one of these three photons artificially, via setting the
corresponding matrix elements of the Liouvillean to zero,
the peaks at~$\tilde\delta=-(\sqrt2-1)$ and~$\tilde\delta=-(\sqrt3-1)/2$
are dramatically decreased. The excitation pathway for the peak 
at~$\tilde\delta=-(\sqrt3-1)/2$ may be understood in a simple way
as follows.
Once again setting~${\cal E}_1=0$ for 
the~$\vert 0\rangle\longleftrightarrow\vert 1 \rangle_-$ transition 
reduces the peak, not only at~$\tilde\delta=-(\sqrt2-1)$, but also
for its close neighbour~$\tilde\delta=-(\sqrt3-1)/2$.

Let us now consider the increase in peak~$v$ of Fig.~\ref{app:3PCR:1}(a). 
This increase is due to a decrease in competition which arises by
eliminating the~$\left\vert 0 \right\rangle\longleftrightarrow\left\vert 1\right
\rangle_-$ transition due to an~$\omega_1$ photon. This competition is
clear from the following analysis. From Fig.~\ref{app:3PCR:1}(d), we
see that peak~$v$ is reduced by eliminating the~$\omega_2$ photon which
drives the~$\left\vert 0 \right\rangle\longleftrightarrow\left\vert 1\right
\rangle_+$ transition. Then Fig.~\ref{app:3PCR:2}(h) shows that peak~$v$
is also reduced by eliminating the~$\omega_2$ 
photon--induced~$\left\vert 1 \right\rangle_+ \longleftrightarrow 
\left\vert 2\right\rangle_+$ transition and, finally, the~$\omega_2$ photon 
is also responsible for the~$\left\vert 2 \right\rangle_+ \longleftrightarrow 
\left\vert 3\right\rangle_+$ transition as shown in Fig.~\ref{app:3PCR:3}(h).
Therefore, removing the~$\omega_1$ photon--induced ~$\left\vert 0 
\right\rangle\longleftrightarrow\left\vert 1\right\rangle_-$ transition
allows a greater population to occur in the~$\left\vert 1 \right\rangle_+$
level in Fig.~\ref{app:3PCR:1}(d) and then, via two more~$\omega_2$--induced
transition, a greater rate~$\left\langle a^{\dag\,3}a^3 \right\rangle$.

In Fig.~\ref{app:3PCR:1}(a) we observe that peak~$vi$ is completely eliminated
by setting~${\cal E}_1=0$ for the~$\left\vert 0 \right\rangle
\longleftrightarrow\left\vert 1\right\rangle_-$ transition. We can see from
Figs.~\ref{app:3PCR:2}(d) and~\ref{app:3PCR:3}(h) that 
an~$\omega_1\longleftrightarrow\omega_2\longleftrightarrow\omega_2$ pathway
exists for~$\left\vert 0\right\rangle \longleftrightarrow
\left\vert 1\right\rangle_-\longleftrightarrow
\left\vert 2\right\rangle_+ \longleftrightarrow
\left\vert 3\right\rangle_+$, which is responsible for peak~$vi$ in the
rate~$\left\langle a^{\dag\,3}a^3 \right\rangle$.
A second, less important, pathway is evident in Figs.~\ref{app:3PCR:2}(b)
and \ref{app:3PCR:3}(d) as the~$\omega_1\longleftrightarrow\omega_2
\longleftrightarrow\omega_2$ 
induced pathway~$\left\vert 0\right\rangle \longleftrightarrow
\left\vert 1\right\rangle_-\longleftrightarrow
\left\vert 2\right\rangle_- \longleftrightarrow
\left\vert 3\right\rangle_+$. Thus, there are two contributing pathways, 
one via~$\left\vert 2\right\rangle_-$ and the other via~$\left\vert 2\right
\rangle_+$, which are both prevented by eliminating the~$\omega_1$
photon--induced~$\left\vert 0 \right\rangle\longleftrightarrow\left\vert 
1\right\rangle_-$ transition. Therefore, the peak in Fig.~\ref{app:3PCR:1}(a)
is reduced for the condition that the~$\vert 0 \rangle\longleftrightarrow
\vert 1\rangle_-$ is suppressed.

We observe in Fig.~\ref{fig:three} that peak~$iv$ is suppressed for
peak~$\tilde\delta=2-\sqrt3$. This involves a resonant excitation 
to~$\vert 1 \rangle_-$ by an~$\omega_1$ photon followed by 
an~$\omega_2$ photon exciting to~$\vert 2 \rangle_-$ with a
detuning of~$(1+\sqrt2-\sqrt3)g_f\doteq 0.68g_f$ followed by an~$\omega_1$
resonant excitation to~$\vert 3\rangle_-$. Another 
dominant pathway involves a resonant excitation to~$\vert 1\rangle_-$
by an~$\omega_1$ photon followed by another~$\omega_1$ photon excitation
to~$\vert 2\rangle_-$ with detuning~$(2-\sqrt2)g_f\doteq 0.59g_f$, followed
by a resonant excitation to~$\vert 3\rangle_-$.

We do not observe a peak at~$\tilde\delta=2-\sqrt3$ because the off-resonant
excitations are small in comparison to excitations near the two 
peaks~$iii$ and~$v$. In contrast, peak~$viii$ also involves a nonresonant
pathway which is similar. This pathway involves a resonant excitation 
to~$\vert 1\rangle_-$
by an~$\omega_1$ photon followed by another~$\omega_1$ photon excitation
to~$\vert 2\rangle_-$ with detuning~$(2-\sqrt2)g_f\doteq 0.59g_f$, followed
by a resonant excitation to~$\vert 3\rangle_+$.
Peak~$viii$ is visible because there is no major peak nearby, but its height
is less than the valley height at~$iv$.

To complete the analysis of 3PCR, we consider the intriguing bump on the graph
in Fig.~\ref{fig:three} 
near~$\tilde\delta\doteq-0.7$. 
There is no vertical line in Fig.~\ref{fig:three}
at this bump because this value of~$\tilde\delta$ does not
correspond to any resonant excitation
pathway. From Figs.~\ref{app:3PCR:1}(b),~\ref{app:3PCR:2}(b) 
and~\ref{app:3PCR:3}(b) we observe that there is a~$3\omega_2$ induced 
excitation pathway along~$\left\vert 0\right\rangle \longleftrightarrow
\left\vert 1\right\rangle_-\longleftrightarrow
\left\vert 2\right\rangle_- \longleftrightarrow
\left\vert 3\right\rangle_-$. 
The frequency for the~$\vert 0\rangle\longleftrightarrow\vert 3\rangle_-$
transition is~$3\omega_2=3\omega-\sqrt3 g_f$ which corresponds 
to~$\omega_2=\omega-g_f/\sqrt3$.
This value of~$\omega_2$ is equivalent to setting~$\tilde\delta=-1/\sqrt3$,
which corresponds to the solid vertical line~$i$ in Fig.~\ref{fig:three}.
The reason for the bump occurring at~$\tilde\delta\doteq -0.7$
rather than at~$\tilde\delta=-1/\sqrt3$ is shown
in Fig.~\ref{fig:multladder} and explained below. 


In Fig.~\ref{fig:multladder} each excitation pathway
involves nonresonant transitions. In each of the three cases,
the transition~$\vert 0\rangle\longleftrightarrow\vert 1\rangle_-$
occurs on resonance ($\tilde\delta=-1$) or slightly off resonance
($\tilde\delta=-1/\sqrt2$ and $\tilde\delta=-1/\sqrt3$).
The~$\tilde\delta=-1$ case does not generate a large population of
the~$\vert 3\rangle_-$ level, because of the significantly off 
resonant excitations to~$\vert 2\rangle_-$ and to~$\vert 3\rangle_-$.
Similarly, the~$\tilde\delta=-1/\sqrt3$ case also does not 
significantly populate~$\vert 3\rangle_-$ because of the large
detuning from~$\vert 2\rangle_-$ compared to the homogeneous 
linewidth~$3\kappa+\gamma/2$. The dominant pathway is 
thus~$\tilde\delta=-1/\sqrt2 \doteq -0.7$, for which the excitation is
close to resonance for~$\vert 1\rangle_-$ and~$\vert 3\rangle_-$
and exactly on resonance for~$\vert 2\rangle_-$: hence the
observed bump at~$\tilde\delta\doteq-0.7$.

In Figs.~\ref{app:3PCR:3}(c) and~(d) we observe that the bump at 
$\tilde\delta\doteq-0.7$ is reduced
by competition with the~$\left\vert 2 \right\rangle_-\longleftrightarrow
\left\vert 3\right\rangle_+$ pathway. By setting either~${\cal E}_1=0$ 
or~${\cal E}_2=0$ for this transition, the bump is increased. Thus, 
excitation along the~$\left\vert 0\right\rangle \longleftrightarrow
\left\vert 1\right\rangle_-\longleftrightarrow
\left\vert 2\right\rangle_- \longleftrightarrow
\left\vert 3\right\rangle_+$ pathway diminishes the 3PCR, and this decrease
is more pronounced for~$3\omega_2 > 3\omega-\sqrt3 g_f$. Therefore, 
a consequence of this competition 
between excitation pathways to~$\vert 3\rangle_-$ and~$\vert 3\rangle_+$
is an apparent shift of the peak
responsible for the bump to lower~$\tilde\delta$. 

Figures \ref{app:3PCR:1}--\ref{app:3PCR:3} provide
sufficient information to extend the analysis above and identify
specific excitation pathways to determine which contributions 
dominate the 3PCR graph of Fig.~\ref{fig:three}. The
cases discussed above are illustrative examples. 
Also a similar analysis can be applied to understand the structure of the
3PCR~$\langle \sigma_+ a^{\dag\,2} a^2 \sigma_-\rangle$, but we do not
perform this analysis here.

\subsection{Difference two--photon count rate}
\label{subsec:difftwoph}

The background subtraction discussed in 
Section~\ref{sec:Master:Twophoton} is relevant to multiphoton contributions.  
The difference--2PCR is constructed in the same way as before,
except that the multi--PCR is taken into account:
\begin{equation}
\Delta^{(n)}(\delta, {\cal E}_1) = w^{(n)}(\delta,{\cal E}_1)
	- w^{(n)}(\delta,{\cal E}_1=0).
\end{equation}
The 2PCR contains contributions from all higher--order terms;
thus, when one of the two chromatic components in the 
bichromatic driving field is shut off, the difference--PCR 
includes subtraction of all higher--order correlation functions.

It is convenient and reasonable to neglect higher--order
multiphoton contributions to the 2PCR and retain only
the term of interest $w^{(2)}$ and the lowest--order 
correction term $w^{(3)}$.
Thus, the 2PCR in the perturbative limit can be approximated by
$w^{(2)} + w^{(3)}$.


The difference--2PCR is depicted in Fig.~\ref{fig:multp}
as a plot of ${\Delta^{(2)}}$ versus $\tilde{\delta}$, including the
3PCR contribution.
After background subtraction, we observe
2PCR peaks for~$\tilde\delta=1-\sqrt2$, $\tilde\delta=\sqrt2-1$ 
and~$\tilde\delta=\sqrt2+1$, depicted as peaks II, III, and V, 
respectively. These three peaks have been identified and
discussed in detail in 
Carmichael {\em et~al.}~1996 and in Sanders {\em et~al.}~1997.
The important
point in Fig.~\ref{fig:multp} is that the 2PCR without 3PCR corrections,
in
Sanders {\em et~al.}~1997,
corresponds closely to the corrected version here. 
That is, the solid and dotted lines in Fig.~\ref{fig:multp} are 
quite similar.
Therefore, 3PCR would not destroy the desired peak in 2PCR. However, 
three--photon effects are manifested on the 2PCR peak in Fig.\ 
\ref{fig:multp}, and this extra structure is of interest. This 
noticeable extra
structure occurs for values of~$\tilde\delta$ corresponding to the 
vertical lines I, IV and VI.

The extra structure due to 3PCR is not surprising. 
This is because 2PCR requires excitation
to~$\vert 2 \rangle_{\pm}$ with zero detuning in order to  observe the best
possible 2PCR. This condition of excitation 
to~$\vert 2\rangle_{\pm}$ is also the prerequisite for
observing 3PCR, as resonant excitation to~$\vert 2\rangle_{\pm}$
ensures a nonnegligible population of~$\vert 3\rangle_{\pm}$.

A particularly significant 3PCR effect is the enhancement of the peak 
at~$\tilde\delta=-(\sqrt2-1)$ and a slight shift in the peak position. 
In Fig.~\ref{fig:multp} we can compute the 2PCR with and without the 
3PCR correction. The 3PCR correction term is significant. This 3PCR
contribution applies for values of~$\tilde\delta$ which contribute to 
peaks~$ii$ and~$iii$ in Fig.~\ref{fig:three}. Although Fig.\ 
\ref{fig:three} applies only for~$g=g_f$, this represents the dominant
contribution after background subtraction. Therefore, Fig.~\ref{fig:three}
is a useful guide for understanding the enhancement of peak II in 
Fig.~\ref{fig:multp}. 

Minor peaks occur at~$\tilde\delta$ values for vertical lines IV and VI in
Fig.~\ref{fig:multp}. The origin of the peak IV is the same
pathway responsible for peak $vi$ in Fig.~\ref{fig:three}. Similarly, peak
VI corresponds to peak $viii$ in Fig.~\ref{fig:three}. The peaks in 
Fig.~\ref{fig:three} enable us to understand the extra structure of the
2PCR in Fig.~\ref{fig:multp}. A quantum trajectory simulation of the
2PCR, which included the 3PCR corrections, multiple atom
effects and finite window times, is shown in Fig.\ 5 of 
Carmichael {\em et~al.}~1996.
Peak IV is slightly discernible in Fig.\ 5 of 
Carmichael {\em et~al.}~1996
although subject to a small shift in the peak position 
(shifts in peak position have been discussed in 
Horvath and Sanders 2001.
The information in Subsection~\ref{subsec:3PCR} enables a complete
understanding of the structure in Fig.~\ref{fig:multp} and, in practice,
the enhanced 2PCR in Fig.~5 of 
Carmichael {\em et~al.}~1996.

\section{Non--simultaneity of photodetections}
\label{sec:nonsimultaneity}

In the following analysis we investigate the appropriate window time for 
counting photon pairs.
The higher--order correlations can be ignored in determining the
optimal window time~$\tau_{\rm w}$ as the correction for this
optimal time will be small.
Photon pair emissions,
which arise due to de--excitation from the~$\vert 2\rangle_{\pm}$ level to
the ground state of the JC ladder,
 are not emitted from the cavity simultaneously due to the
randomness of photoemission due to the cavity linewidth.
The detection of a photon pair thus depends on identifying a window
time~$\tau_{\rm w}$ such that,
for two photons detected with temporal separation~$t < \tau_{\rm 
w}$,
the two photons are deemed to be members of a pair.
If~$t > \tau_{\rm w}$, the photons
are deemed to be independent single photons (not members of a pair). 
Here we determine the optimal window time $\tau_{\rm opt}$ which maximises the 
counting rate of genuine pairs relative to the rate of false pair counts. 

\subsection{The two--photon count rate} 
\label{subsec:2PCRn} 
 
The 2PCR can be obtained in more than one way.
Ideally one would have a perfectly efficient photodetector that
detects all photons leaving one cavity mirror.
The photodetector would then provide a complete record of photon
emissions from the cavity as a function of~$t$.
A perfect coincidence then arises as two simultaneously 
detected photons at some time~$t$.
However, there are two challenges.
One challenge is that there does not exist a perfectly efficient photodetector.
Therefore, some pairs of photons are observed as single-photon emissions
because one member of the pair escapes observation.
In fact some pairs are missed altogether because both photons escape
detection.
The other challenge concerns the detection of two simultaneously created 
photons.
Although created simultaneously, the emission from the cavity is
not simultaneous due to the randomness of the emission time
resulting from the nonzero cavity lifetime~$\kappa^{-1}$.

We consider two photons to be coincident provided that they
arrive within the window time interval~$\tau_{\rm w}$.
The choice of window time is not obvious, and it is our aim here
to determine what the window time should be.
As the two simultaneous photons can be separated by a time
of order $\kappa^{-1}$,
as discussed above,
the window time~$\tau_{\rm w}$ might be expected to be on the 
order of $\kappa^{-1}$.
However, our purpose here is to consider the choice of $\tau_{\rm w}$
in detail and to identify the optimal choice of window time~$\tau_{\rm w}$
that will produce the strongest 2PCR.

The choice of optimal window time is further complicated by the
method of detecting nearly simultaneous photons.
In the ideal case discussed above of a perfect photodetector
yielding a record of all photon emissions from the cavity,
one can then define a two--photon event as taking place if
a second photon arrives between times~$t_0$ and $t_0 + \tau_{\rm w}$,
{\em conditioned} on a photodetection at time~$t_0$.
We refer to this rate as the {\em conditional} difference--2PCR and
define this rate to be
\begin{equation}
\label{2PCR:con}
\Delta^{(2)}_{\rm con}(\delta, {\cal E}_1,\tau_{\rm w})
	\equiv
	\lim_{t_0 \rightarrow \infty}\frac{1}{\tau_{\rm w}}
	\int_{t_0}^{t_0+\tau_{\rm w}} dt \,
	\Delta^{(2)}(t,\delta,{\cal E}_1).
\end{equation}
for~$\Delta^{(2)}(t,\delta,{\cal E}_1)=w^{(2)}(t,\delta,{\cal E}_1)
-w^{(2)}(t,\delta,{\cal E}_1=0)$ with~$w^{(2)}(t,\delta,{\cal E}_1)$
given by Eq.~(\ref{eq:w2n}).
(This conditional difference--2PCR for a window time
$\tau_{\rm w} = \kappa^{-1}$
was used in the quantum trajectory analysis of PCS in 
Carmichael {\em et~al.}~1996).

Another natural way to measure the 2PCR is by counting all photon pairs
defined as being separated by an interval less than~$\tau_{\rm w}$.
This 2PCR is referred to as the {\em unconditional} difference--2PCR and does
not rely on starting the count for the second photon conditioned
on detecting the first photon.
The definition of the unconditional 2PCR is
\begin{eqnarray} 
\label{2PCR:unc} 
\Delta^{(2)}_{\rm unc}(\delta,{\cal E}_1,\tau_{\rm w})
	&=& \lim_{t_0 \rightarrow \infty}
	\frac{2}{\tau_{\rm w}^{2}} 
	\int_{t_0}^{t_0+\tau_{\rm w}}  dt^\prime
	\int_{t_0}^{t^\prime} dt\,
	w^{(2)}(t^\prime-t,\delta,{\cal E}_1). 
\end{eqnarray}
As shown in~\ref{app:a}, this expression can be simplified to read
\begin{equation} 
\label{two:result} 
\Delta^{(2)}_{\rm unc}(\delta, {\cal E}_1,\tau_{\rm w})
	= \frac{2}{\tau_{\rm w}^{2}}  
	\int_0^{\tau_{\rm w}} du  \int_0^u d\varsigma\,
	w^{(2)} (\varsigma,\delta, {\cal E}_1).
\end{equation} 

Both the conditional and unconditional difference--2PCR involve measuring
photons and inferring whether two photons may be considered to have been
`simultaneously emitted'.  This determination is based on whether two
photons arrive within a specified time interval.  In the conditional case,
two photons are treated as being members of the same `simultaneously
emitted' pair if the second follows the first by less than the time
window~$\tau_{\rm w}$; in the unconditional case, they are considered to
be members of a pair if the temporal separation of the photodetections is
less than the time window~$\tau_{\rm w}$.  Both definitions are reasonable
and depend on the nature of the detection scheme.  The window time for the
conditional difference--2PCR will always be less than for the
unconditional difference--2PCR.  The conditional process is more
efficient, in a sense, because the detection interval is always triggered
by the first photon, whereas the latter case allows part of an interval to
pass even before the first photon is detected. Thus, in the unconditional
case, a longer interval is required to be able to detect the first photon
and then subsequently wait for the second photodetection.

We provide analytical solutions for the two extreme cases 
in~\ref{app:a}.
The window time can be extremely long
$(\kappa \tau_{\rm w} \gg 1)$,
yielding expression (\ref{long:time}),
or extremely short $(\kappa \tau_{\rm w} \ll 1)$,
yielding expression (\ref{short:time}) for both conditional and
unconditional 2PCR.
The short window time ($\tau_{\rm w} \longrightarrow$ 0) 
was the basis of the analysis of 2PCS in 
Sanders {\em et~al.}~1997.
In this treatment both the conditional and unconditional 2PCR at time~$t$
is approximated by $\langle:\hat{n}^2(t):\rangle$. In the long--time limit
the 2PCR is dominated by Poissonian statistics.

The choice of optimal window time~$\tau_{\rm opt}$ depends on the
technique for observing two--photon coincidences, but another factor
must also be considered.
The purpose of 2PCS is to observe two--photon decay resonances
from the combined atom-cavity system.
These 2PCR peaks are shown in Fig.~\ref{fig:Fig1}
as a function of the normalized scanning field frequency~$\tilde\delta$.
The choice of $\tau_{\rm w}$ will depend on which peak is
being observed.
However, the best peak for observing a two--photon decay resonance
occurs for~$\tilde\delta = 1+\sqrt2$.

\subsection{The Peak--to--Valley Ratio (PVR)}
\label{subsec:PVR}

\begin{figure}[2]
\begin{picture}(200,380)(45,10)
\font\gnuplot=cmr10 at 10pt
\gnuplot
\rotatebox{90}{\resizebox{450pt}{530pt}{\includegraphics{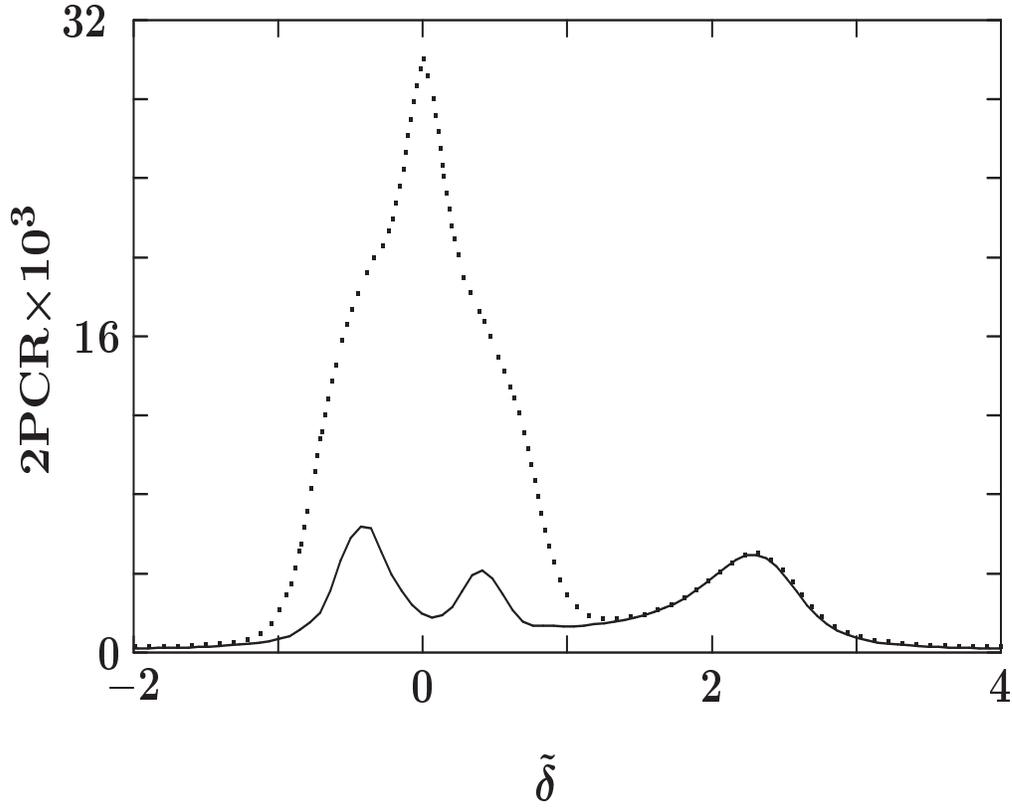}}}
\end{picture}
\caption{The 2PCR vs the normalized detuning~$\tilde\delta$ with  
${\cal E}_1/\kappa=0.5$, ${\cal E}_2/\kappa=0.5$, $g_f/\kappa=9$, 
$\gamma/\kappa=2$ for the inhomogeneously broadened system: $w^{(2)}(\delta,{\cal E}_1)$ is the solid line
and~$\Delta^{(2)}(\delta,{\cal E}_1)$ is the dashed line.
}
\label{fig:Fig1}
\end{figure}

\begin{figure}[3]
\begin{picture}(200,380)(45,10)
\font\gnuplot=cmr10 at 10pt
\gnuplot
\rotatebox{90}{\resizebox{450pt}{530pt}{\includegraphics{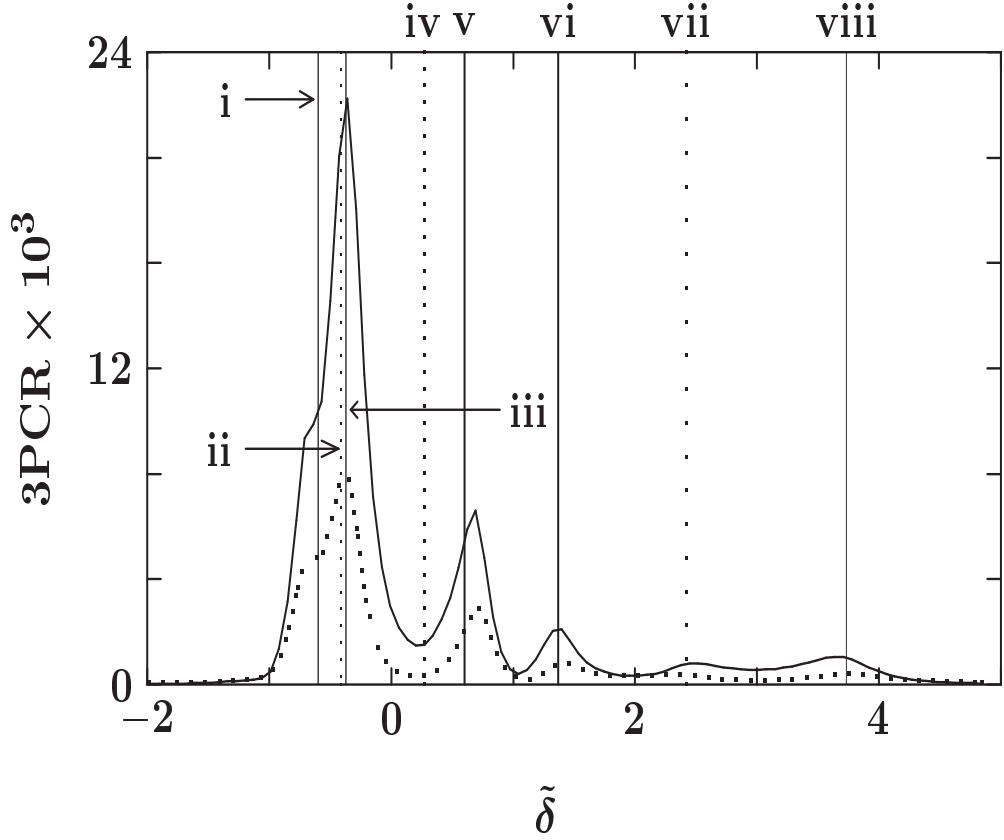}}}
\end{picture}
\caption{Three--photon count rate (3PCR) vs normalized 
	scanning 
	frequency for~$g=g_f=9\kappa$, ${\cal E}_1/\kappa=1/\sqrt2$,
	${\cal E}_2/\kappa=\sqrt{2}$ and $\gamma/\kappa=2$.
	The solid line applies to~$\langle a^{\dag\,3}a^3\rangle$,
	and the dotted line is 
	for~$\langle \sigma_+ a^{\dag\,2}a^2\sigma_-\rangle$. 
	Alternating solid and dotted vertical lines are placed at values
	of~$\tilde\delta$ for which the 3PCR peaks are expected, 
	namely,~$\tilde\delta\in \{-1/\sqrt3,-(\sqrt2-1),-(\sqrt3-1)/2,
	2-\sqrt3,1/\sqrt3,(\sqrt3+1)/2,\sqrt2+1,2+\sqrt3 \}$}
\label{fig:three}
\end{figure}

\begin{figure}[4]
\begin{picture}(200,350)(70,30)
\font\gnuplot=cmr10 at 10pt
\gnuplot
\rotatebox{90}{\resizebox{400pt}{500pt}{\includegraphics{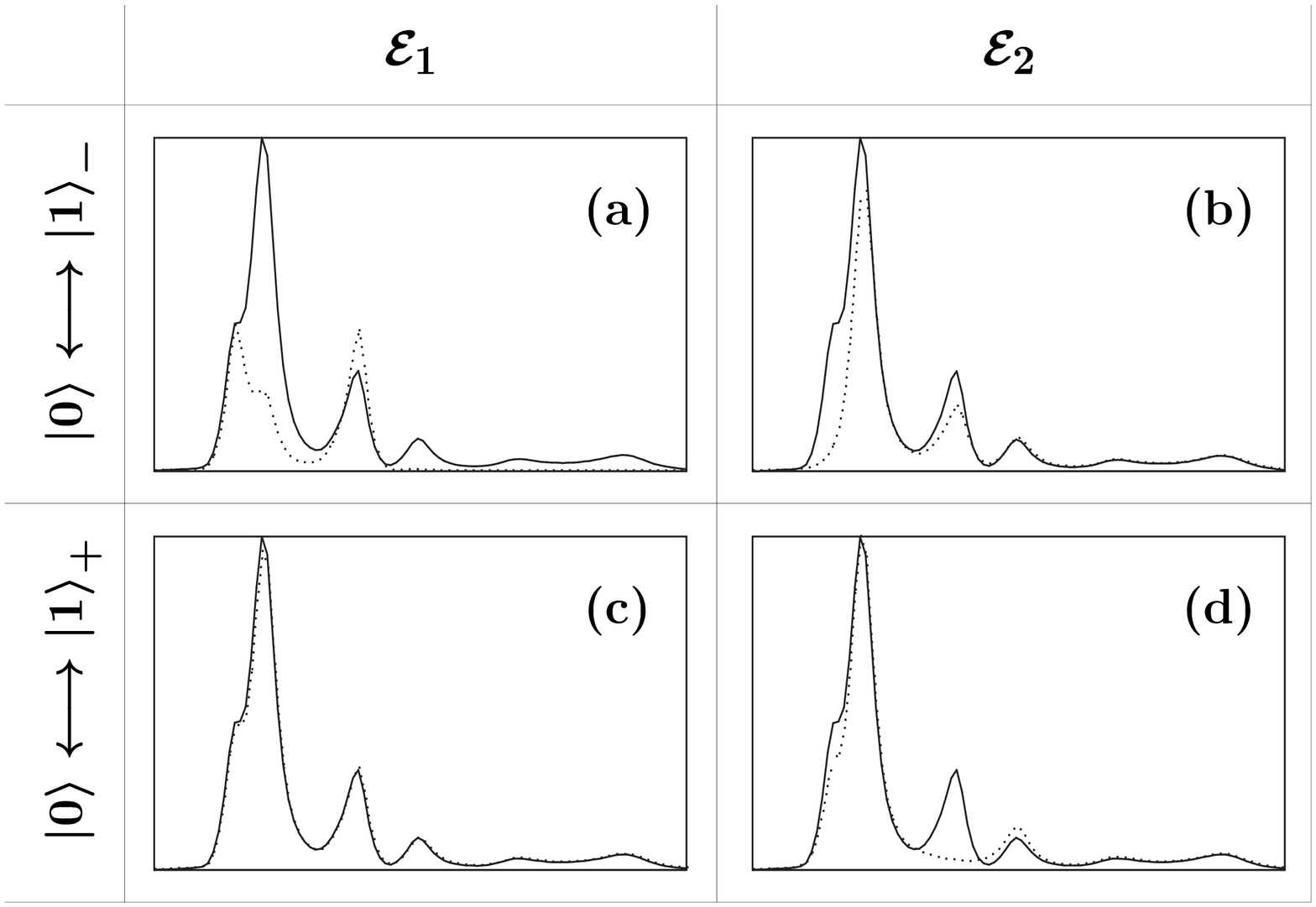}}}
\end{picture}
\caption{The rate~$\left \langle a^{\dag\,3}a^3\right\rangle$,
which is depicted in Fig.~\ref{fig:three}, is reproduced here as a
solid line in each of the four figures (a),(b),(c),(d).
The rate~$\left \langle a^{\dag\,3}a^3\right\rangle$ is repeated
as a dotted line with~${\cal E}_1=0$ for 
the~$\vert 0\rangle\longleftrightarrow\vert 1\rangle_-$ transition
in (a), with~${\cal E}_2=0$ 
for~$\vert 0\rangle\longleftrightarrow\vert 1\rangle_-$ in (b),
with~${\cal E}_1=0$ for 
the~$\vert 0\rangle\longleftrightarrow\vert 1\rangle_+$ transition
in (c) and with~${\cal E}_2=0$ for 
the~$\vert 0\rangle\longleftrightarrow\vert 1\rangle_+$ transition in (d).
We present the four figures by allocating the two rows to each of the
two transitions
~$\vert 0\rangle\longleftrightarrow\vert 1\rangle_{\pm}$ and the 
columns to whether~${\cal E}_1=0$ or~${\cal E}_2=0$ for the
transition in question.
	}
\label{app:3PCR:1} 
\end{figure}

\begin{figure}[5]
\begin{picture}(0,450)(-10,10)
\font\gnuplot=cmr10 at 10pt 
\gnuplot
\resizebox{700pt}{800pt}{\includegraphics{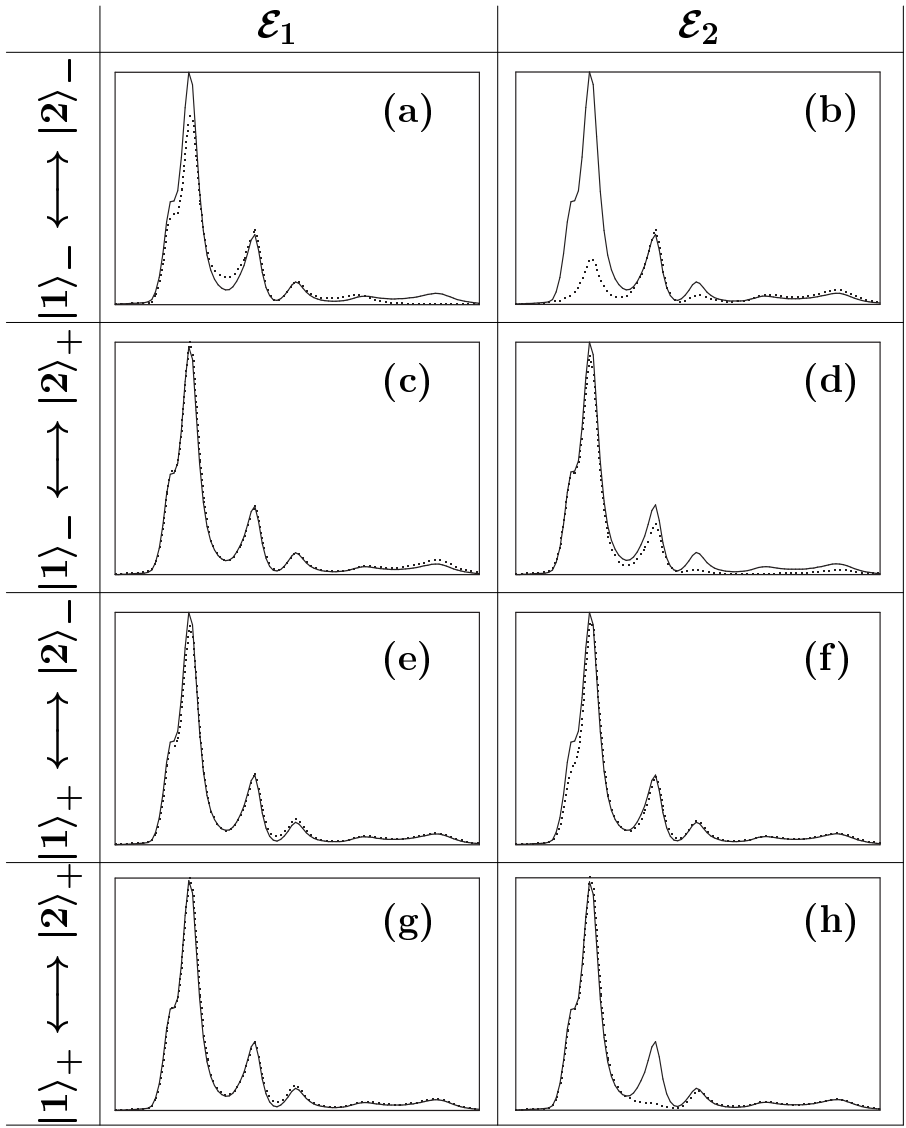}}
\end{picture}
\caption{As for Fig.~\ref{app:3PCR:1}, with the solid line depicting the
rate~$\left \langle a^{\dag\,3}a^3\right\rangle$ and the dotted line
corresponding to the same rate but with 
either~${\cal E}_1=0$ (first column) or~${\cal E}_2=0$ (second column)
for the four 
transitions~$\vert 1\rangle_{\varepsilon}
\longleftrightarrow\vert 2\rangle_{\varepsilon}'$ with 
~$\varepsilon,\, \varepsilon'\in \{-,+\}$.}
\label{app:3PCR:2} 
\end{figure}

\begin{figure}[6]
\begin{picture}(0,450)(-10,10)
\font\gnuplot=cmr10 at 10pt
\gnuplot
\resizebox{700pt}{800pt}{\includegraphics{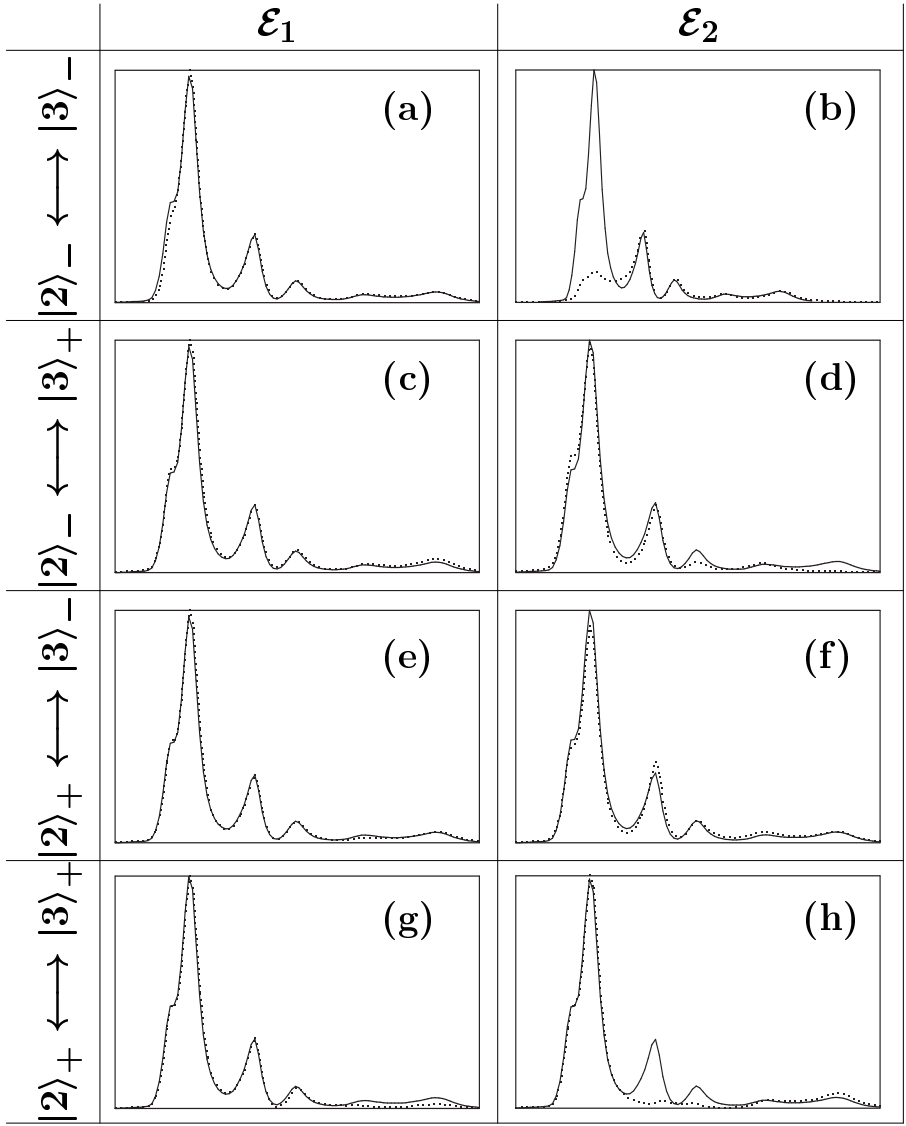}}
\end{picture}
\caption{As for Fig.~\ref{app:3PCR:2} but with~${\cal E}_1=0$ 
(first column) or~${\cal E}_2=0$ (second column)
for the four 
transitions~$\vert 2\rangle_{\varepsilon}
\longleftrightarrow\vert 3\rangle_{\varepsilon}'$ with 
~$\varepsilon,\, \varepsilon'\in \{-,+\}$.}
\label{app:3PCR:3} 
\end{figure}

\begin{figure}[7]
\begin{picture}(0,350)(0,10)
\font\gnuplot=cmr10 at 10pt 
\gnuplot
\resizebox{350pt}{350pt}{\includegraphics{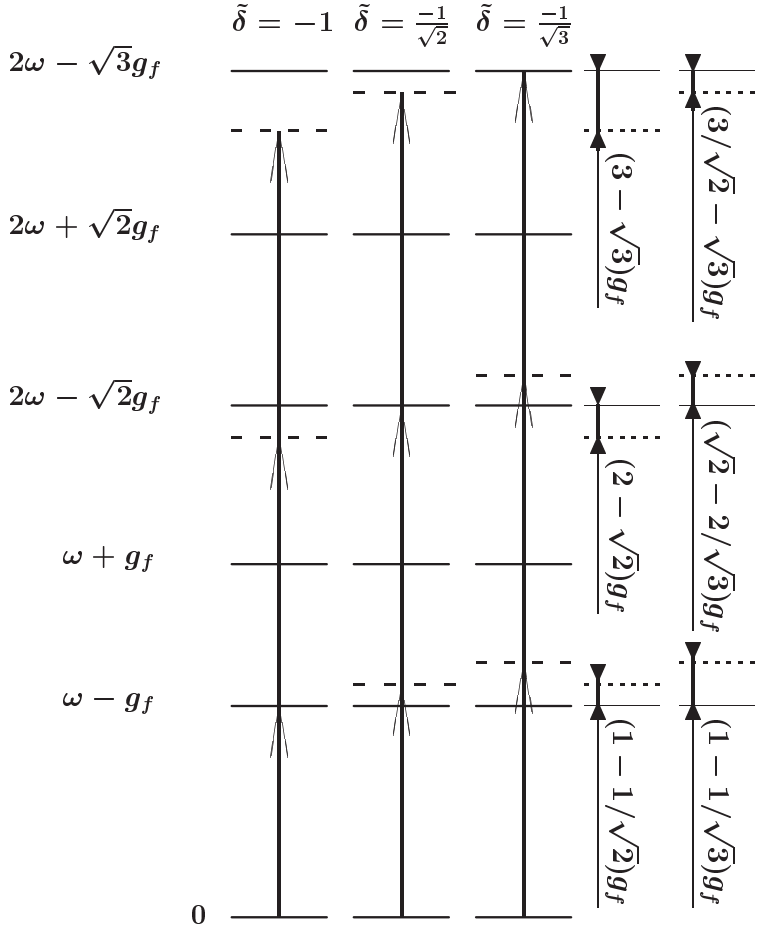}}
\end{picture}
\caption{Excitations to~$\vert 3\rangle_-$ for~$\tilde\delta=-1$, 
~$\tilde\delta=-1/\sqrt2$ and for~$\tilde\delta=-1/\sqrt3$ are presented
here. For~$\tilde\delta=-1$, resonant excitation occurs 
to~$\vert 1 \rangle_-$,
followed by off--resonant excitation to~$\vert 2\rangle_-$, with a 
detuning of~$(2-\sqrt2)g_f\doteq0.59g_f$, followed by off-resonant
excitation to~$\vert 3\rangle_-$ with a 
detuning of~$(3-\sqrt3)g_f\doteq 1.3 g_f$. 
 For~$\tilde\delta=-1/\sqrt2$,
off--resonant excitation from~$\vert 0 \rangle$ to~$\vert 1\rangle_-$,
with a detuning of~$(1-1/\sqrt2)g_f\doteq 0.29g_f$, is followed by an
resonant excitation to~$\vert 2\rangle_-$, followed by an  
off--resonant excitation to~$\vert 3\rangle_-$ with a
detuning of~$(3/\sqrt2-\sqrt3)g_f\doteq0.39g_f$.
For~$\tilde\delta=-1/\sqrt3$,
off--resonant excitation from~$\vert 0 \rangle$ to~$\vert 1\rangle_-$,
with a detuning~$(1-1/\sqrt3)g_f\doteq 0.42g_f$, is followed by an
off-resonant excitation to~$\vert 2\rangle_-$ with a
detuning of~$(\sqrt2-2/\sqrt3)g_f\doteq 0.26g_f$, followed by a 
resonant excitation to~$\vert 3\rangle_-$.}
\label{fig:multladder}
\end{figure}

\begin{figure}[8]
\begin{picture}(200,380)(45,10)
\font\gnuplot=cmr10 at 10pt
\gnuplot
\rotatebox{90}{\resizebox{450pt}{530pt}{\includegraphics{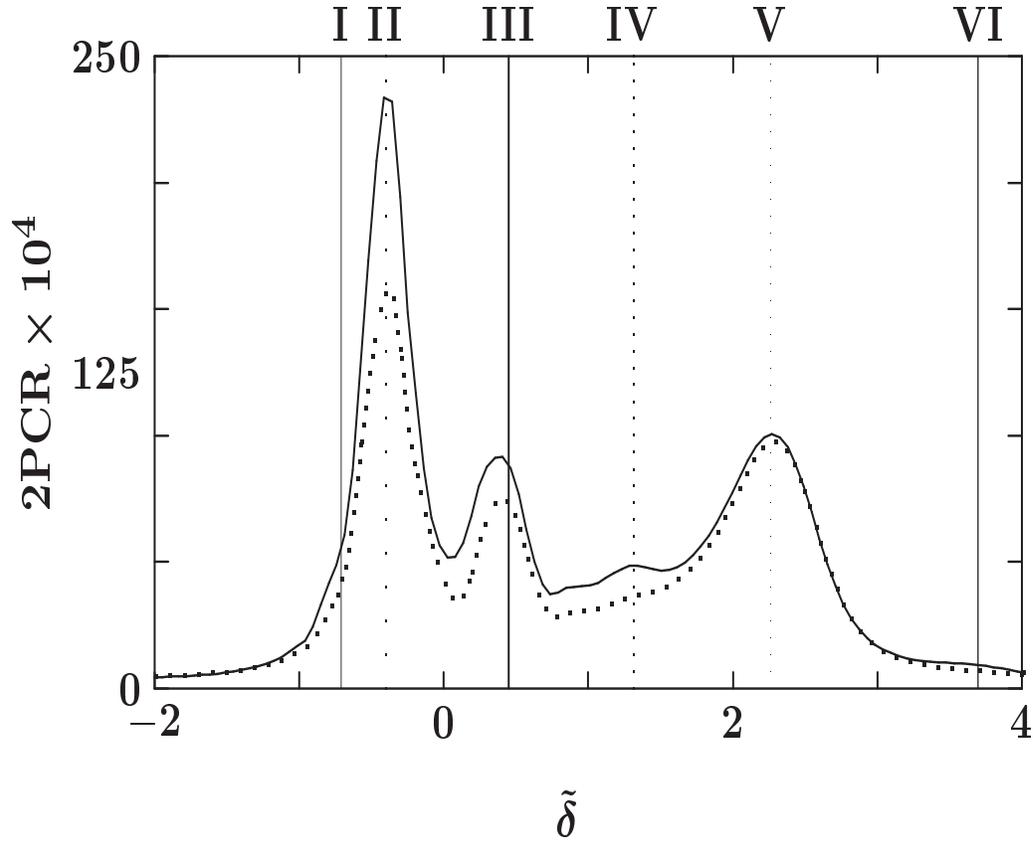}}}
\end{picture}
\caption{Difference--2PCR
	vs normalized 
	scanning frequency for an inhomogeneously broadened system
	and~${\cal E}_1/\kappa=1/\sqrt2$,~${\cal E}_2/\kappa=\sqrt2$,
	$\gamma/\kappa=2$. The solid line corresponds 
	to~$\Delta^{(2)}+\Delta^{(3)}$ and the dotted line to~$\Delta^{(2)}$.}
\label{fig:multp}
\end{figure}

In Fig.~\ref{fig:Fig1} the 2PCR peak at~$\tilde\delta=1+\sqrt2$ resides over
a background 2PCR
that is largely independent of~$\tilde\delta$.
In this Section we employ the parameters~${\cal E}_1={\cal E}_2=
\kappa/2$ (equality of~${\cal E}_1$ and~${\cal E}_2$ helps to
reduce the non--resonant two--photon absorption background as in
Sanders {\em et~al.}~1997)
whereas, in 
Section~\ref{sec:multiphoton},~${\cal E}_1={\cal E}_2/2=
\kappa/\sqrt2$ (as used in 
Carmichael {\em et~al.}~1996).
In this Section we 
choose lower driving field strengths which has the effect
of producing smaller peak shifts than those studied in 
Sec.~\ref{sec:multiphoton}. For this choice of parameters, 
the 2PCR background 
is~$\overline{\langle : n^2 : \rangle} \approx 2.1\times 10^{-5}$.
Let us characterise the quality of a 2PCR peak by the ratio of the
peak height to the height of the background 2PCR.
We can understand this figure of merit in terms of the ratio of signal 
to noise,
where signal is the 2PCR from the desired two--photon decay events,
and the background noise corresponds to two--photon decays arising
from unwanted off-resonance two-quantum excitations and decay events.
The peak-to-valley ratio (PVR) is determined by the height of the
peak to the height of the background (or valley) 2PCR.
The optimal window time
$\tau_{\rm w} = \tau_{\rm opt}$
is defined such that the PVR
for this 2PCR is maximal.
That is, either a larger or a smaller choice of the window time would
reduce the value of the PVR, thereby making the peak more difficult to discern
from background events.

There are other concerns besides the PVR in choosing the window time.
For example, choosing a much shorter window time could improve the
PVR but also lengthen the run time of the experiment required to 
accumulate enough signal.
In other words, the absolute height of the peak is also a matter of concern in 
determining the feasibility of the experiment and is determined by the
allowable timescale of the experiment. The minimum height would need to
be on the order of~$T^{-1}$ for~$T$ the timescale of the data collection.

The PVR is obtained numerically. The matrix continued--fraction method is used
to solve the master equation to determine the peak height in
Sanders {\em et~al.}~1997.
The background, or
valley, is solved analytically by assuming a large detuning for
the scanning field. The details are provided in appendix~\ref{app:b}. 

\begin{figure}[9]
\begin{picture}(200,580)(-100,-310)
\font\gnuplot=cmr10 at 10pt
\gnuplot
\resizebox{360pt}{280pt}{\includegraphics{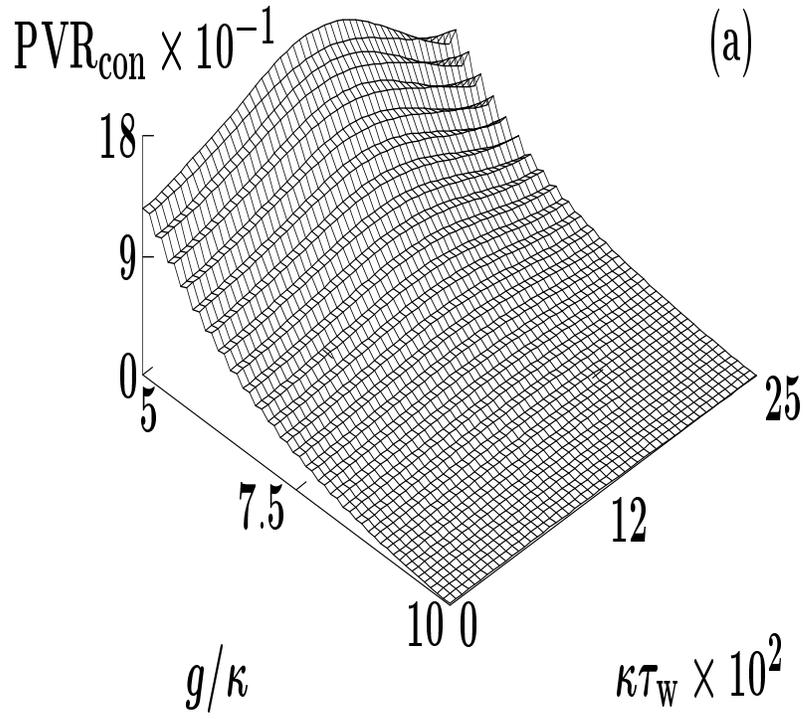}}
\end{picture}
\begin{picture}(200,200)(100,20)
\font\gnuplot=cmr10 at 10pt
\gnuplot
\resizebox{360pt}{280pt}{\includegraphics{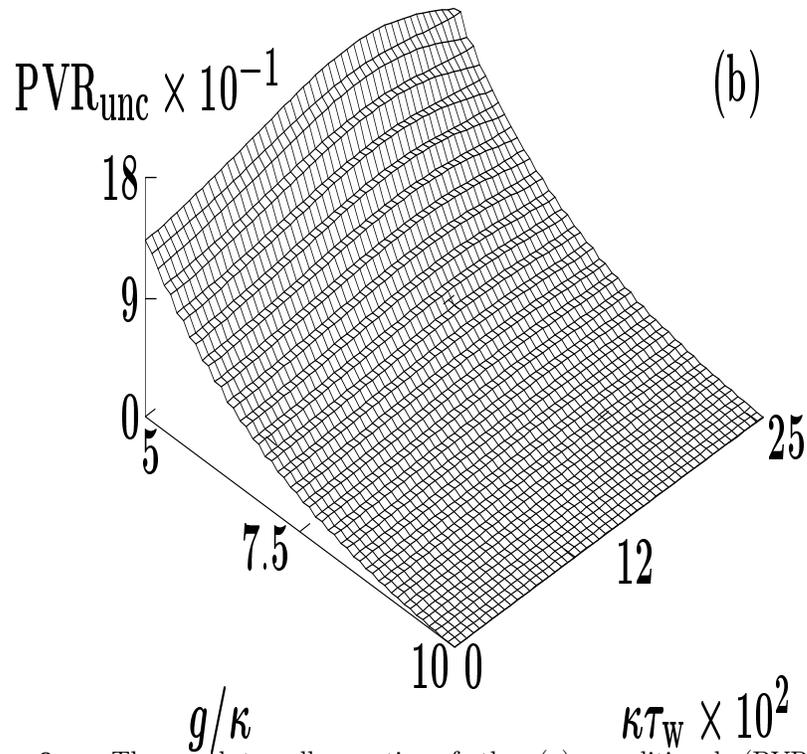}}
\end{picture}
\caption{The peak-to-valley ratio 
	of the (a)~conditional (PVR$_{\rm con}$) and 
	(b)~unconditional (PVR$_{\rm unc})$ 2PCR over the scaled
	coupling strength~$g/\kappa$ and the scaled window 
	time~$\kappa\tau_{\rm w}$
	for the scaled loss rate~$\gamma/\kappa=2$.}
	\label{fig:surfPVR}
\end{figure}

The 2PCR for large~$\tilde\delta$ is given
by Expressions~(\ref{offres:c2PCS}) and~(\ref{offres:2PCS}).
The PVR for 2PCR is thus
\begin{equation}
\label{PVR}
{\rm PVR_{\xi}} = \frac{ \Delta^{(2)}_{\xi} (\tilde\delta,{\cal E}_1, 
\tau_{\rm w}) }
	{ \Delta^{(2)}_{{\rm 0}\,\xi} (\tau_{\rm w} ) }
\end{equation}
with~$\xi\in\{\rm con, \rm unc\}$.
This convenient notation enables us to discuss both conditional and
unconditional 2PCR with a single compact notation.
In Fig.~\ref{fig:surfPVR} surface plots of the 
PVR vs~$g$ 
and~$\tau_{\rm w}$ reveal that the PVR increases as
$g$ decreases.
This decrease is due to the background 
2PCR (for~$\tilde\delta$ large) becoming small as shown in 
Fig.~\ref{fig:Fig1}, whereas the pair signal rate does not decline
as quickly for decreasing~$g$.
The increase in PVR is due to the fact that the resonant
frequency for the transition
$|0\rangle\longleftrightarrow |1\rangle_-$ is~$\omega-g$, for a particular
subensemble with coupling strength~$g$, 
whereas the pump field frequency is constrained 
to satisfy~$\omega_1=\omega-g_f$.
Hence, as~$g$ 
decreases, the pump field drives the system more and more off resonance.
Thus, the likelihood of driving the system from~$\vert 0\rangle$ 
to~$\vert 1\rangle_-$, followed by a photoemission, then repeating
the driving~$|0\rangle\longleftrightarrow |1\rangle_-$ and a second
photoemission, decreases rapidly with diminishing~$g$: hence the background
2PCR due to rapid driving to~$\vert 1\rangle_-$ from~$\vert 0\rangle$ is
quite small. A lesser contribution to the 2PCR, due to~$2\omega_1$ photons
causing off--resonant 
excitation~$\vert 0\rangle \longleftrightarrow \vert 2\rangle_-$, also 
diminishes rapidly for decreasing~$g$ because the excitation moves
progressively further from the resonance condition. Although the 2PCR due
to de--excitation from~$\vert 2\rangle_{\pm}$ decreases with 
diminishing~$g$, the
2PCR for the background falls quicker thereby increasing the PVR.

\begin{figure}[10]
\begin{picture}(200,580)(0,-250)
\font\gnuplot=cmr10 at 10pt
\gnuplot
\rotatebox{90}{\resizebox{350pt}{530pt}{\includegraphics{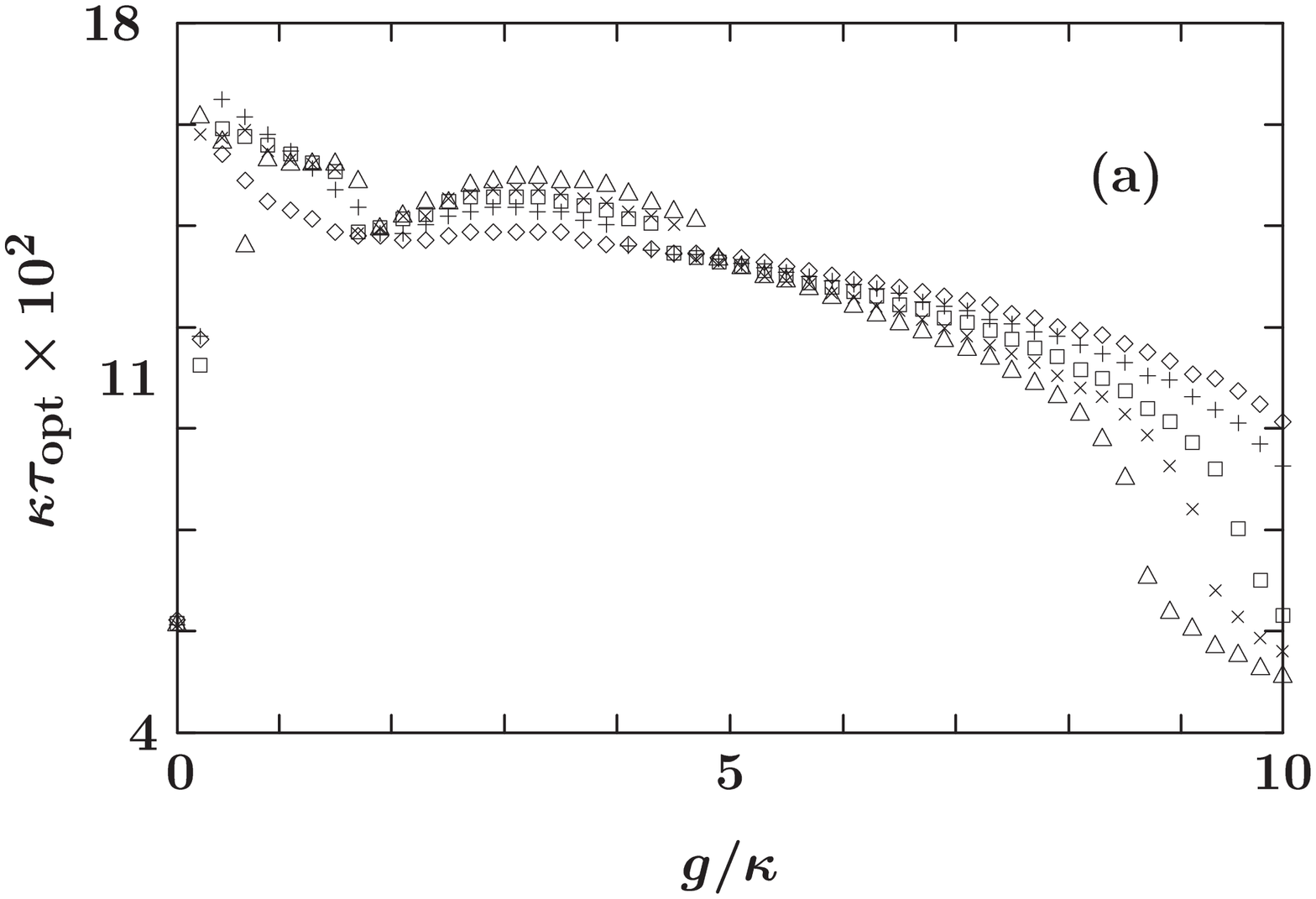}}}
\end{picture}
\begin{picture}(300,580)(190,20)
\font\gnuplot=cmr10 at 10pt
\gnuplot
\rotatebox{90}{\resizebox{350pt}{515pt}{\includegraphics{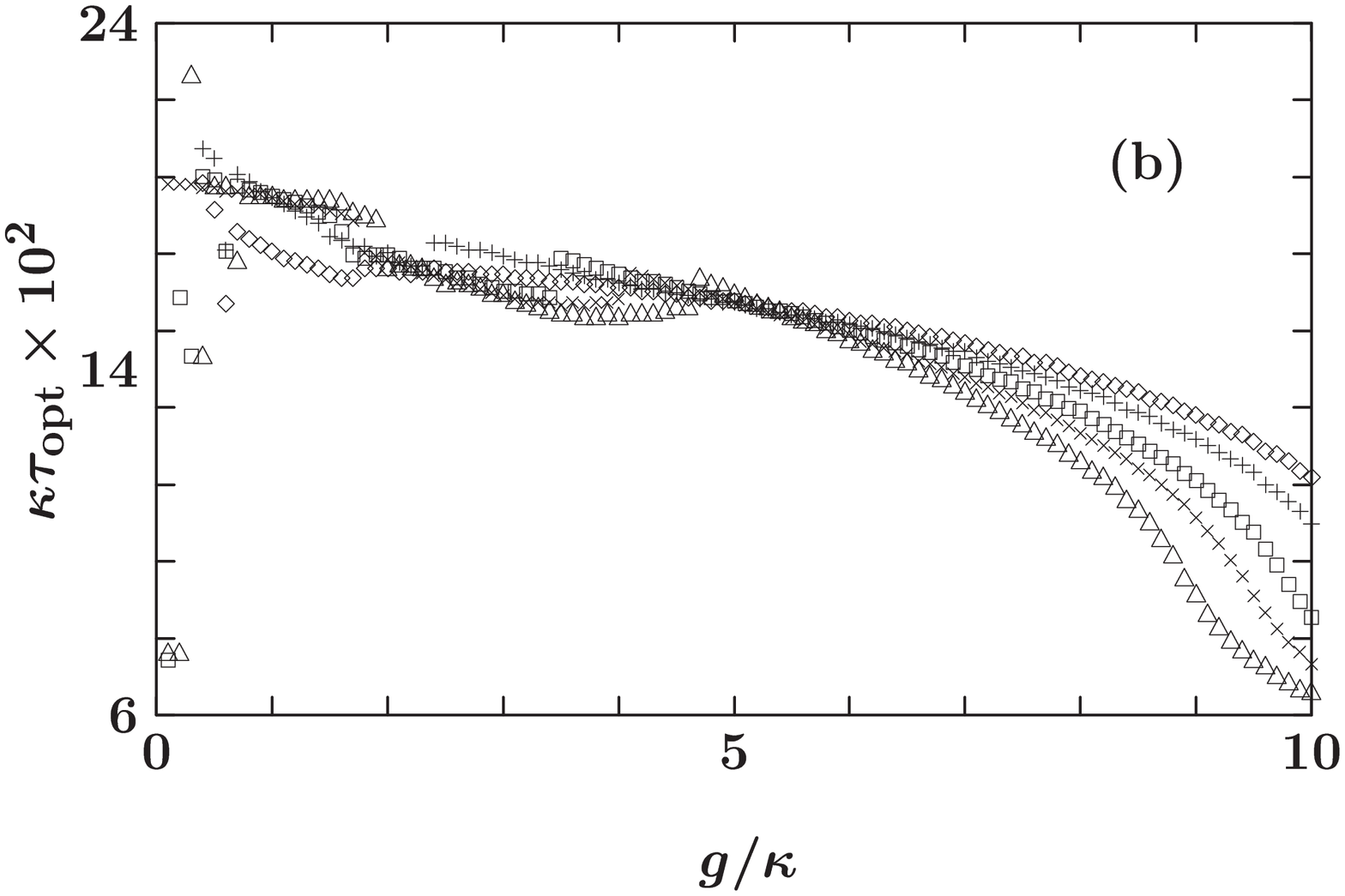}}}
\end{picture}
\caption{The scaled optimal window time $\kappa \tau_{\rm opt}$ 
	vs the scaled coupling strength $g/\kappa$ for
	$\diamond$ $\gamma/\kappa=0.2$, $+$ $\gamma/\kappa=2$,
	$\Box$ $\gamma/\kappa = 5$,
	$\times$ $\gamma/\kappa=7$ and
	$\triangle$ $\gamma/\kappa=10$ for (a)~the conditional 
	and~(b) the unconditional 2PCR.}
	\label{fig:optvsg}
\end{figure}

As our objective is to optimise the PVR, we choose the value of~$\tau_{\rm w}$
which, for each~$g$, maximises the PVR. A single peak for each~$g$ is evident
in Figs.~\ref{fig:surfPVR}(a) and~(b). In Fig.~\ref{fig:optvsg}
we plot the value of the dimensionless quantity~$\kappa\tau_{\rm w}$
vs~$g$ which maximises the PVR. This choice of~$\tau_{\rm w}$ would 
provide the optimal PVR for 2PCS if~$g$ could be fixed. It is clear from 
Fig.~\ref{fig:optvsg} that 
the window time~$\tau_{\rm w}$ for achieving the optimal PVR is
an order of magnitude smaller than $\kappa^{-1}$. The rapid decline 
of~$\tau_{\rm opt}$ in the vicinity of~$g=g_f=9\kappa$ is due to the
driving from~$\vert 0\rangle$ to~$\vert 1\rangle_-$, followed by
a photoemission, and repeating the process (as discussed above).
Thus~$\tau_{\rm w}$ must be substantially reduced for~$g\approx g_f$
in order to excise the single--photon contributions to the measured 2PCR.

Fig.~\ref{fig:optvsg} presents~$\kappa\tau_{\rm opt}$ vs~$g/\kappa$ for
a range of values~$0.2\leq \gamma/\kappa\leq 10.0$. The effect of
varying~$\gamma/\kappa$ is to change the homogeneous linewidth
particularly for the~$\vert 1\rangle_-$ level, which is important
for the spurious contribution to the 2PCR, as discussed above. The rapid
decline in~$\tau_{\rm opt}$, as~$g$ approaches~$g_f$, varies with the choice
of~$\gamma$. For~$g$ small, the excitation from~$\vert 0\rangle$ 
to~$\vert 1\rangle_-$ by an~$\omega_1$ photon is sufficiently detuned
that varying the homogeneous linewidth does not have much of the effect,
hence the approximately flat curve for~$\kappa \tau_{\rm opt}$ 
vs~$g/\kappa$ in Fig.~\ref{fig:optvsg}.
However, the point of decline for~$\kappa\tau_{\rm opt}$ does depend
on the homogeneous linewidth, and we observe that the decline occurs for
smaller~$g$ with~$\gamma/\kappa$ large, and the decline occurs for 
large~$g$ as~$\gamma/\kappa$ decreases. Fig.~\ref{fig:optvsg}
supports the earlier suggestion that the rapid oscillation 
between~$\vert 0\rangle$ and~$\vert 1\rangle_-$, with concomitant
photoemissions, is indeed responsible for diminishing the PVR.

\begin{figure}[11]
\begin{picture}(200,380)(45,10)
\font\gnuplot=cmr10 at 10pt
\gnuplot
\rotatebox{90}{\resizebox{450pt}{550pt}{\includegraphics{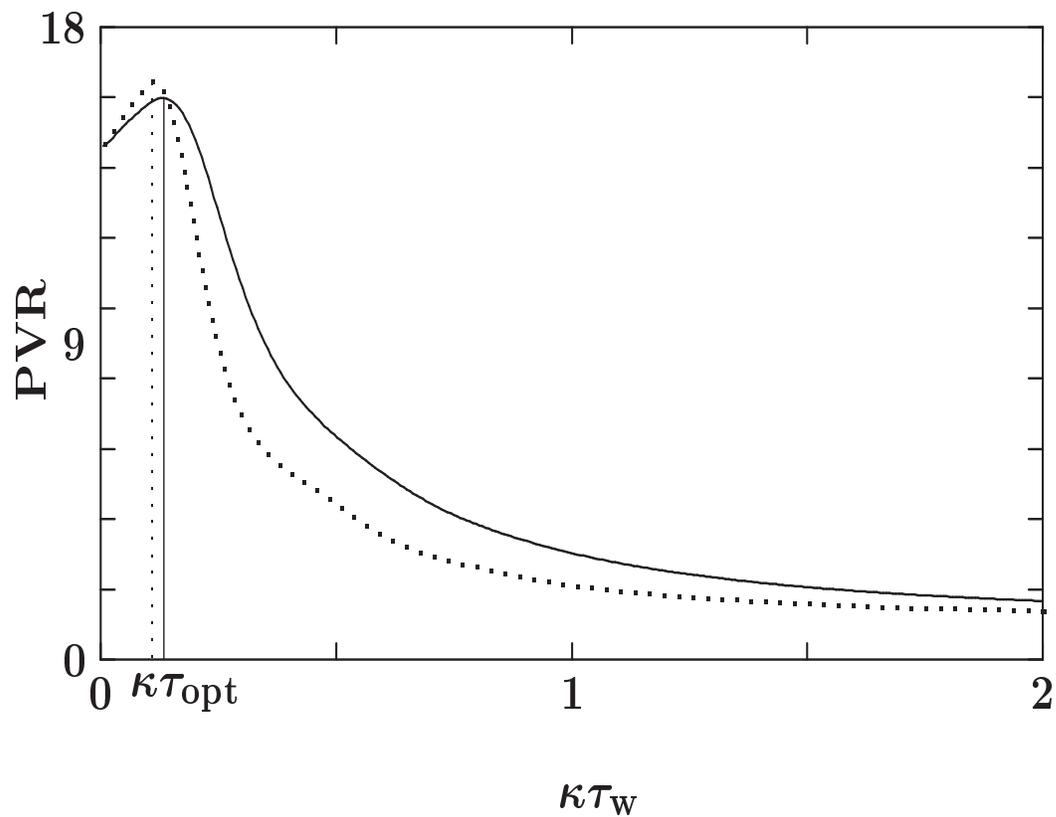}}}
\end{picture}
\caption{The  peak-to-valley ratio (PVR) of the conditional (solid line) 
	and unconditional (dashed line) 2PCR 
	(for the masked atomic beam) vs 
	the scaled window time~$\kappa\tau_{\rm w}$
	for the scaled loss rate $\gamma/\kappa=2$.
	}  
\label{fig:avgPVR} 
\end{figure}

Fig.~\ref{fig:optvsg} is useful for understanding~$\tau_{\rm opt}$,
but the experimentally relevant choice of~$\tau_{\rm opt}$ is determined
by taking the trace with respect to~$\overline{\rho}$ in 
Eq.~(\ref{density:matrix}). Using the same~$P(g)$ as throughout this
paper, we obtain the PVR for the density matrix~$\overline{\rho}$.
The highest PVR is plotted against~$\kappa\tau_{\rm opt}$ in 
Fig.~\ref{fig:avgPVR}. Fig.~\ref{fig:avgPVR} depicts quite clearly the 
optimal choice~$\tau_{\rm opt}$ which maximises the PVR, and, 
for~$\gamma/\kappa=2$, we observe that~$\kappa\tau_{\rm opt}=0.111$
for conditional difference--2PCR and~$\kappa\tau_{\rm opt}=0.135$
for unconditional difference--2PCR.

\begin{figure}[12]
\begin{picture}(200,380)(45,10)
\font\gnuplot=cmr10 at 10pt
\gnuplot
\rotatebox{90}{\resizebox{450pt}{550pt}{\includegraphics{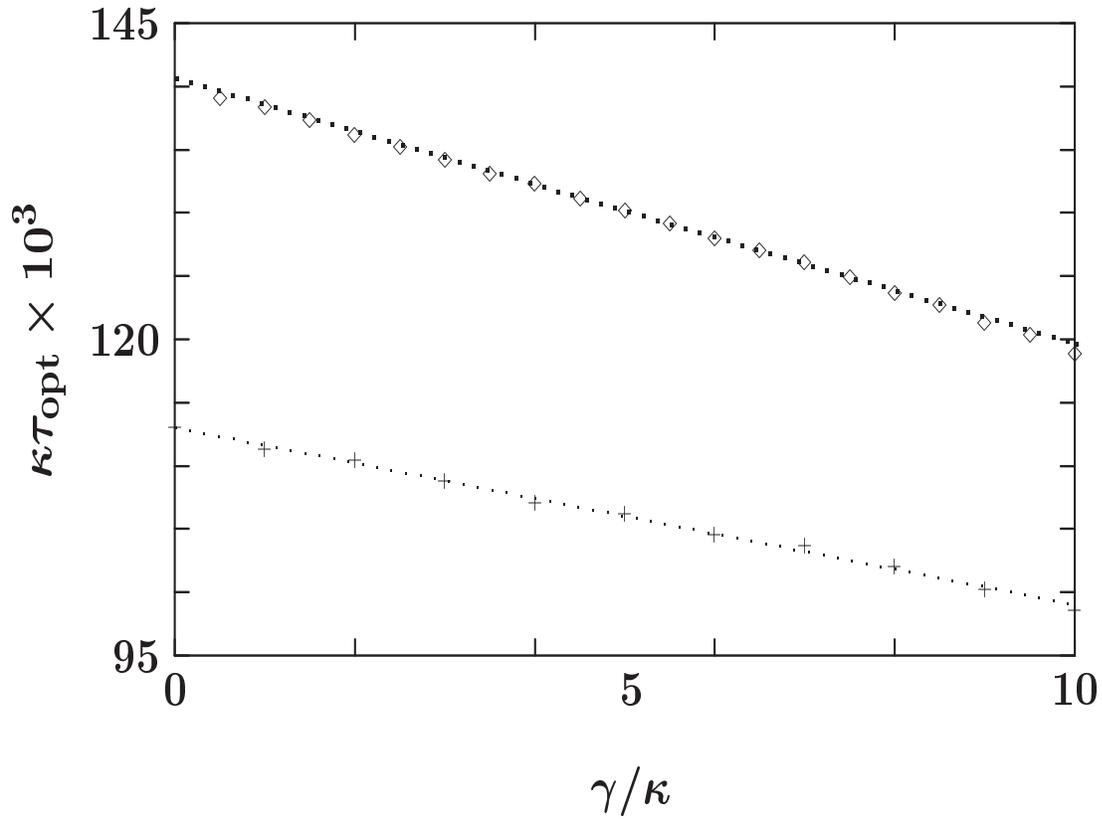}}}
\end{picture}
\caption{The scaled optimal window time 
	$\kappa\tau_{\rm opt}$ vs the scaled loss rate~$\gamma/\kappa$  
	for the masked atomic beam with inhomogeneous broadening. 
	The symbol $+$ corresponds to the
	conditional~$\tau_{\rm opt}$ and~$\diamond$ to the 
	unconditional~$\tau_{\rm opt}$. Linear regression methods yield the
	two lines. For the conditional case, the slope is~$-1.4\times10^{-3}$, 
	the intercept is~$0.11$, and the correlation coefficient 
	is~$r=-0.9983$. 
	For the unconditional case, the slope is~$-2.1\times10^{-3}$, 
	the intercept is~$0.14$, and the correlation coefficient 
	is~$r=-0.9995$.   
	}
\label{fig:resopt} 
\end{figure}

Using this approach we plot~$\kappa\tau_{\rm opt}$ vs~$\gamma/\kappa$ in 
Fig.~\ref{fig:resopt}. As~$P(g)$ is heavily weighted in favor of low~$g$,
the high--$g$ decline in Fig.~\ref{fig:optvsg} is not so important.
Hence, we observe that~$\tau_{\rm opt}$ is relatively insensitive to the
choice of~$\gamma$, and the dependence is linear with correlation
coefficients of~$-0.9983$ and~$-0.9995$
for the conditional and unconditional cases respectively. 
The dependence of~$\tau_{\rm opt}$ on~$\gamma$
is given by~$\kappa\tau_{\rm opt}=-1.4\times10^{-3}\gamma/\kappa+0.11$
for the conditional difference--2PCR 
and $\kappa\tau_{\rm opt}=-2.1\times10^{-3}\gamma/\kappa+0.14$
for the unconditional difference--2PCR. The
variation of~$\kappa\tau_{\rm opt}$ for the range of~$\gamma/\kappa$ is
very small. It is clear that~$\tau_{\rm opt}$ is an order of magnitude
smaller than~$\kappa^{-1}$, and the dominant deleterious contribution
to the difference--2PCR is the 
cycling~$\vert 0\rangle\longleftrightarrow\vert 1\rangle_-$ with
corresponding photoemissions.

\section{Conclusions}
\label{sec:conclusion}

Photon coincidence spectroscopy
(Carmichael {\em et~al.}~1996, Sanders {\em et~al.}~1997,
Horvath {\em et~al.}~1999, Horvath and Sanders 2001)
is a
promising technique for observing unambiguous experimental signatures of
quantum field effects in quantum electrodynamics. This 
quantum field signature consists of
two--photon spectral peaks at certain locations (with respect to detuning
between the two chromatic components of the driving field). However,
the peaks are not located exactly as predicted in previous 
studies due to multiphoton 
contributions to the two--photon count 
rate, due to non--simultaneity of the photon pairs emitted from the
cavity and due to the effect of the cavity lifetime.

We have considered these effects here. The multiphoton contributions are 
responsible for additional structure in the two--photon count rate peaks
as shown in Fig.~\ref{fig:multp}. We observe that the higher--order 
contributions enhance the peak heights, rather than diminish them. 
Although additional structure is predicted by Fig.~\ref{fig:three},
for the chosen parameters, the additional structure is negligible on the
two--photon count rate peaks. The technique of 2PCS is robust against
multi--photon contributions to the two--photon count rate.

Another important quantity is the detection window time~$\tau_{\rm w}$.
This quantity is the duration over which two photons are treated
as though they originate from the same intracavity photon pair. For a
separation of greater than~$\tau_{\rm w}$, the photons are regarded as
having been created independently. We consider two detection protocols,
conditional and unconditional difference--2PCR. In each case the
optimal window time is an order of magnitude smaller than the cavity
lifetime due primarily to 
the need to reduce the~$\vert 0\rangle\longleftrightarrow
\vert 1\rangle_-$ contribution to the two--photon count rate. We also note
the negligible effect of varying the inhibited spontaneous emission 
rate~$\gamma$ for the coupling strength distribution~$P(g)$ which is heavily
weighted in favour of low~$g$.

These effects, namely multiphoton contributions to the pair count rate and
the optimal choice of window time, are important in the design of
photon coincidence spectroscopy schemes and in interpreting the results.
Photon coincidence spectroscopy has been shown here to be quite robust,
even with these potential deleterious effects being included in
the simulations.
 
\section*{Acknowledgments}
This project has been funded by an Australian Research Council Large
Grant, an Australian Research Council Small Grant, a Macquarie University
Research Grant and the Macquarie University Postgraduate Research Fund.
We acknowledge the valuable assistance provided by B.~Wielinga in the
early stages of this research, and valuable discussions with H.~J.~Carmichael
throughout this undertaking.

\clearpage

\clearpage

\clearpage

\clearpage

\clearpage

\clearpage

\clearpage

\clearpage

\clearpage

\clearpage

\clearpage

\clearpage


\appendix

\section{Conditional and unconditional two--photon count rate}
\label{app:a} 
 
In the long--time limit,
the conditional two--photon count rate (2PCR) is given by
\begin{eqnarray}
\label{two:photo:b}
\Delta^{(2)}_{\rm con} (\delta, {\cal E}_1,\tau_{\rm w}) 
= \frac{1}{\tau_{\rm w}} (2\kappa)^{2} \int_{0}^{\tau_{\rm w}} dt
\left \langle : \hat{n}(0) \hat{n}(t): \right \rangle
\end{eqnarray}
with the trace being taken over~$\overline{\rho}$ in 
Eq.~(\ref{density:matrix}).
If the time window $\tau_{\rm w}$ is large,
compared to $\kappa^{-1}$ (the cavity lifetime), the two photons are highly 
decorrelated, and we can approximate
\begin{equation} 
	\label{long:time:exp} 
	\left\langle : \hat{n}(0) \hat{n}(t) : \right\rangle
	\longrightarrow \left\langle \hat{n}(0) \right\rangle^{2}. 
\end{equation} 
Thus,
\begin{equation} 
\label{long:time} 
\Delta^{(2)}_{\rm con}(\delta,{\cal E}_1 \tau_{\rm w}) \longrightarrow 
	(2\kappa)^2\left\langle \hat{n}(0) \right\rangle^{2}. 
\end{equation}
This count rate reflects the Poissonian nature of the count statistics for 
long window times. 
On the other hand, for $ \kappa \tau_{\rm w} \ll 1$, 
the correlation between photon pairs cannot be neglected. Hence, 
the count rate reduces to 
\begin{eqnarray}	 
\label{short:time} 
\Delta^{(2)}_{\rm con}(\delta, {\cal E}_1,\tau_{\rm w})\longrightarrow 
	(2\kappa)^2\left \langle : \hat{n}^2(0) : \right\rangle, 
\end{eqnarray} 
which is the approximation employed in 
Sanders {\em et~al.}~1997.

Similarly, in the long-time limit, the unconditional 2PCR 
is
\begin{equation} 
\label{two:photo:a} 
\Delta^{(2)}_{\rm unc}\left(\delta,{\cal E}_1,\tau_{\rm w} \right)
	= \frac{2}{\tau_{\rm w}^{2}} (2\kappa)^{2}
	\int_0^{\tau_{\rm w}} dt^\prime 
	\int_0^{t^\prime}dt 
	\left\langle : \hat{n}(t)\hat{n}(t^\prime) : \right \rangle.
\end{equation} 
This expression can be simplified as we show below.
First we make the substitution~$u_{\pm} = (t^\prime \pm t)/\sqrt{2}$.
We also introduce the notation~$d^2 u=du_- du_+$ and let~${\cal V}$ 
be the union
of the two regions~$\{0 < u_- < \tau_{\rm w}/\sqrt2, 0 < u_+ < u_-\}$ 
and~$\{\tau_{\rm w}/\sqrt2 < u_- < \sqrt2 \tau_{\rm w},
0 < u_+ < \sqrt2\tau_w-u_-\}$

This substitution transforms the above double integral into 
the sum of two double integrals:
\begin{eqnarray} 
\label{two:arrange} 
	\Delta^{(2)}_{\rm unc}(\delta, {\cal E}_1, &\tau_{\rm w}&) =  
	\frac{2}{\tau_{\rm w}^{2}} (2\kappa)^{2}  
	\int\!\int_{\cal V} d^2 u
	\left\langle : \hat{n} \left(u_-'\right) 
	\hat{n} \left(u_+'\right) :
	\right\rangle \nonumber \\
	& = & \frac{2}{\tau_{\rm w}^{2}} (2\kappa)^{2} 
	\int\!\int_{\cal V} d^2 u
	\left \langle : \hat{n}(0) 
	\hat{n}(\sqrt{2}u_{+}) : \right\rangle
\end{eqnarray}
for~$u_{\pm}'=(u_+\pm u_-)/\sqrt{2}$.
The advantage of this expression is that the two-time photon
number correlation depends on only one term in the double 
integral instead of both terms in the double integral.

Greater simplification is possible and desirable for studying the
short and long window time~$\tau_{\rm w}$.
Substituting $u_{\pm}= w_{\pm}/\sqrt{2}$ transforms 
Eq.~(\ref{two:arrange}) to
\begin{eqnarray} 
\label{two:arrange:one} 
\Delta^{(2)}_{\rm unc}(\delta,{\cal E}_1,\tau_{\rm w}) & = & 
	\frac{1}{\tau_{\rm w}^{2}} (2\kappa)^{2}\Bigg[ 
	\int_0^{\tau_{\rm w}} dw_- 
	\int_0^{w_-} dw_+
		\nonumber	\\ &&+
	\int_{\tau_{\rm w}}^{2\tau_{\rm w}} dw_- 
	\int_0^{2\tau_{\rm w}-w_-} dw_+ \Bigg] 
	\left\langle : \hat{n}(0) \hat{n} (w_+) : \right\rangle,
\end{eqnarray} 
which reduces to 
\begin{eqnarray} 
\label{two:res} 
\Delta^{(2)}_{\rm unc}(g,\delta,\tau_{\rm w}) & = &
	\frac{2}{\tau_{\rm w}^{2}} (2\kappa)^{2} \int_0^{\tau_{\rm w}} du 
        \int_0^{u} dw \nonumber \\ 
	& & \times
	\left\langle : \hat{n}(0) \hat{n} (w) : \right\rangle.
\end{eqnarray}

For large ($\tau_{\rm w} \gg \kappa^{-1}$) and small 
($\tau_{\rm w} \ll \kappa^{-1}$) window times~$\Delta^{(2)}_{\rm unc}$ reduces 
identically to~$\Delta^{(2)}_{\rm con}$ as shown 
in equations~(\ref{long:time}) and (\ref{short:time}).

\section{Background of conditional and unconditional 2PCR}
\label{app:b}

For the scanning field far off resonance ($\delta$ large),
the time-dependent component of the Liouvillean~${\cal L}$
can be ignored in the rotating picture.
Thus, in the rotating picture, 
the master equation can be written as~$ \dot{\rho} = {\cal L} \rho $
with $\cal L$ time-independent.
The coupling strength~$g$ is fixed quantity, and averaging over~$P(g)$
(the inhomogeneous broadening case) is not considered in this appendix.
If $\rho$ is expressed as a vector, then~$\cal L$ can be expressed 
as a complex matrix with~$ \{ - \lambda_n | n \in {\cal Z}_{N^2} \} $
the set of eigenvalues for~$N$
the number of levels in the Jaynes-Cummings ladder retained after truncation,
and~Re$(\lambda_n)\geq 0$.
The density matrix can be approximated by the sum
\begin{equation}
\label{rhosum}
{\rho}(t) = \sum_{n=1}^{N^2}  
{\rho}_n e^{- \lambda_n (t-t_0)}
\end{equation}
for~$ \{ \rho_n \} $ a set of time-independent $N \times N$ matrices. Thus, 
the conditional 2PCR~(\ref{2PCR:con}) can be written as 
\begin{eqnarray} 
\label{offres:c2PCS} 
\Delta^{(2)}_{\rm 0\,con}(\tau_{\rm w}) &=&   c_0 + \frac{1}{\tau_{\rm w}}  
	\int_0^{\tau_{\rm w}} dt 	
	\times  \sum_{n=1}^{N^2} c_n 
	\exp{\left[ -\lambda_n t \right]} \nonumber \\
	&=&  \! c_0 \!
	\!  + \! \sum_{n=1}^{N^2} \! \frac{ c_n } { \mu_n }\!
	\left \{ \! 1- e^{ -\mu_n  \!} \right \},
\end{eqnarray}
with~$\mu_n\equiv\lambda_n\tau_{\rm w}$ and~$\lambda_n\neq 0$ for~$n>0$.
Here the subscript~$0$ on the left--hand side of Eq.~(\ref{offres:c2PCS})
is used to designate that we are considering the case of
a monochromatic driving field.
In the long window time limit,
\begin{equation}
\label{limit:Delta}
c_0  = \lim_{\tau_{\rm w} \longrightarrow \infty}  
\Delta^{(2)}_{\rm 0\,con}  
(\tau_{\rm w}). 
\end{equation}  
Expansion~(\ref{offres:c2PCS}) provides a useful method for calculating 
$\Delta^{(2)}_{\rm 0\,con}(\tau_{\rm w})$ by diagonalising the
Liouvillean superoperator for the monochromatically--driven case. 
The function $\Delta^{(2)}_{\rm 0\,con}(\tau_{\rm w})$ 
is monotonically increasing because 
$\partial \Delta^{(2)}_{\rm 0\,con}
/\partial \tau_{\rm w} > 0$ if $\tau_{\rm w} \longrightarrow 
\infty$.   
Thus, $\partial \Delta^{(2)}_{\rm 0\,con}
/\partial \tau_{\rm w}
\longrightarrow 0$ as the function approaches the limit given by 
Eq.~(\ref{limit:Delta}).
 
In the same way, the unconditional 2PCR can be obtained: 
\begin{eqnarray} 
\label{offres:2PCS} 
\Delta^{(2)}_{\rm 0\,unc}(\tau_{\rm w}) 
	 &=&   c_0 + \frac{2}{\tau_{\rm w}^{2}}   
	\int_0^{\tau_{\rm w}}    
	du \int_0^{u} dw  \nonumber \\
	&\times& \sum_{n=1}^{N^2} c_n \exp{\left[ -\lambda_n w \right]} \nonumber \\
	&=& c_0	
	 + \! 2 \! \sum_{n=1}^{N^2} \frac{ c_n } { \mu_n }\!
	\left\{ \! \frac{ e^{ -\mu_n } - 1}{ \mu_n }+ 1 \! \right\}\! . 
\end{eqnarray}  
In the~$\tau_{\rm w}\longrightarrow\infty$  
limit,~$\Delta^{(2)}_{0\,{\rm unc}}(\tau_{\rm w})\longrightarrow c_0$. 
This is the same value for the background difference--2PCR as obtained
for the conditional 2PCR. In summary, the Liouvillean superoperator is
diagonalised. The master equation is solved in the
~$\tau_{\rm w}\longrightarrow\infty$ limit to 
obtain~$\rho_{\rm ss}=\rho(t\longrightarrow\infty)$.
Then~$c_0=(2\kappa)^2{\rm Tr}(a^{\dag\,2}a^2\rho_{\rm ss})$. 
This quantity is the
background 2PCR for both the conditional and unconditional difference--2PCRs. 

\vspace{0.5cm}

\section*{References}
	Carmichael H J, Kochan P and Sanders B C 1996 {\em Phys.\ Rev.\
	Lett.\ } {\bf 77} 631--4	\\
	Hood C J, Lynn T W, Doherty A C, Parkins A S and Kimble H J
	2000 {\em Science} {\bf 287}~1447 \\ 
	Horvath L and Sanders B C 2001 {\em Phys.\ Rev.}~A
	{\bf 63} 053812 \\
	(-----2001 {\em Preprint} quant-ph/011079) \\
	Horvath L, Sanders B C and Wielinga B F 1999 {\em J.\ Opt B:
	Quantum and Semiclassical.~Opt.\ } {\bf 1} 446--51 \\
	Jaynes E T and Cummings F W 1963 {\em Proc.\ IEEE} 
	{\bf 51} 89 \\
	Pinkse P W H, Fischer T, Maunz P and Rempe G 2000
	{\em Nature} {\bf 404} 365 \\
	Sanders B C, Carmichael H J and Wielinga B F 1997
	{\em Phys.\ Rev.}~A {\bf 55} 1358--70 \\ 	
	Thompson R J, Turchette Q A, Carnal O and Kimble H J 1998
	{\em Phys.\ Rev.}~A {\bf 57} 3084 \\	
	Turchette Q A, Hood C J, Lange W, Mabuchi H and Kimble H J 1995
	{\em Phys.\ Rev.\ Lett.}~{\bf 75}~4710 \\

\end{document}